\documentclass[a4paper, twoside, reqno, 12pt, dvips]{amsart}
\RequirePackage{fix-cm} 

\usepackage{fixltx2e}     

\usepackage[english]{babel}
\usepackage[latin1]{inputenc}

\usepackage{eucal}
\usepackage{esint}
\usepackage{amsgen}
\usepackage{amsthm}
\usepackage{xspace}
\usepackage{amssymb}
\usepackage{amsmath}
\usepackage{MnSymbol}
\usepackage{amsfonts}
\usepackage{verbatim}
\usepackage{mathrsfs, dsfont}

\usepackage[nice]{nicefrac}

\usepackage{a4wide}

\usepackage{microtype}
\usepackage[bf, up, hang]{caption}

\usepackage{indentfirst}
\usepackage{graphicx}
\usepackage{subfigure}
\usepackage[section]{placeins}
\usepackage{psfrag}

\usepackage{algorithm, algpseudocode} 

\usepackage[usenames, dvipsnames, pdf]{pstricks}
\usepackage{epsfig}
\usepackage{pst-grad} 
\usepackage{pst-plot} 

\usepackage{color}
\definecolor{oneblue}{rgb}{0.0, 0.0, 0.85}
\definecolor{darkgrey}{rgb}{0.273, 0.281, 0.30}

\usepackage{xcolor}
\definecolor{lightgray}{gray}{0.9}

\usepackage[bookmarks, urlcolor=blue, citecolor=blue, linkcolor=blue, pagecolor=blue, colorlinks, hyperfigures]{hyperref}
\hypersetup{hyperindex=true, pdftitle={}, pdfstartview=Fit, pdfhighlight=/P, bookmarksnumbered=false, bookmarksopen=false, colorlinks=true, urlcolor=oneblue, linkcolor=oneblue, citecolor=oneblue, pagecolor=oneblue, plainpages=false, pageanchor=true, pagebackref=true}

\usepackage[compact]{titlesec}

\titleformat{\section}{\normalfont\Large\bfseries\sffamily\center\color{darkgrey}}{\thesection.}{0.5em}{}{}
\titleformat{\subsection}{\normalfont\large\bfseries\sffamily\color{darkgrey}}{\thesubsection.}{0.4em}{}{}
\titleformat{\subsubsection}{\normalfont\normalsize\bfseries\sffamily\color{darkgrey}}{\thesubsubsection.}{0.3em}{}{}

\titlespacing*{\section}{1.0em}{1.0em}{0.8em}[0em]
\titlespacing*{\subsection}{1.0em}{1.0em}{0.8em}[0em]
\titlespacing*{\subsubsection}{1.0em}{0.7em}{0.6em}[0em]

\usepackage{titletoc}

\contentsmargin{0.0em}
\dottedcontents{section}[2.5em]{\addvspace{0.45em}\bfseries}{1.0em}{0pc}
\dottedcontents{subsection}[4.5em]{}{2.6em}{0pc}
\dottedcontents{subsubsection}[5.5em]{}{3.0em}{0pc}

\usepackage{fancyhdr}
\usepackage{lastpage}

\newcommand*\Title{On the Galerkin method for the Serre equations}
\newcommand*\Authors{D.~Mitsotakis, B.~Ilan, D.~Dutykh}

\usepackage{acronym}
\acrodef{cB}{{\bf `classical' Boussinesq}}
\acrodef{FEM}{{\bf Galerkin / Finite-Element Method}}
\acrodef{RK}{{\bf Runge-Kutta}}
\acrodef{DSWs}{{\bf dispersive shock waves}}

\numberwithin{equation}{section}

\newcommand{\ie}{\emph{i.e.}}

\newcommand{\R}{\mathbb{R}} 
\newcommand{\N}{\mathbb{N}} 
\renewcommand{\H}{\mathcal{H}} 
\renewcommand{\O}{\mathcal{O}} 
\newcommand{\BS}{\mathcal{B}_{\rm Serre}} 
\newcommand{\BB}{\mathcal{B}_{\rm cB}} 
\renewcommand{\P}{\mathcal{P}} 
\newcommand{\eps}{\varepsilon}

\newcommand{\h}{\tilde{h}} 
\renewcommand{\u}{\tilde{u}} 
\newcommand{\err}{\mbox{E}} 
\newcommand{\Fex}{F_{\rm exact}} 
\newcommand{\xa}{{x^{\ast}}} 
\newcommand{\ta}{\tau^{\ast}} 

\makeatletter
\newcommand{\sech}{\mathop{\operator@font sech}}
\newcommand{\sgn}{\mathop{\operator@font sgn}}
\newcommand{\cosech}{\mathop{\operator@font cosech}}
\makeatother

\begin{document}

\title[\Title]{On the Galerkin / finite-element method for the Serre equations}

\author[D.~Mitsotakis]{Dimitrios Mitsotakis$^*$}
\address{School of Natural Sciences, University of California, Merced, 5200 North Lake Road, Merced, CA 95343, USA}
\email{dmitsot@gmail.com}
\urladdr{http://dmitsot.googlepages.com}
\thanks{$^*$ Corresponding author}

\author[B.~Ilan]{Boaz Ilan}
\address{School of Natural Sciences, University of California, Merced, 5200 North Lake Road, Merced, CA 95343, USA}
\email{bilan@ucmerced.edu}
\urladdr{http://faculty.ucmerced.edu/bilan}

\author[D.~Dutykh]{Denys Dutykh}
\address{University College Dublin, School of Mathematical Sciences, Belfield, Dublin 4, Ireland \and LAMA, UMR 5127 CNRS, Universit\'e de Savoie, Campus Scientifique, 73376 Le Bourget-du-Lac Cedex, France}
\email{Denys.Dutykh@ucd.ie}
\urladdr{http://www.denys-dutykh.com/}

\begin{abstract}
A highly accurate numerical scheme is presented for the Serre system of partial differential equations, which models the propagation of dispersive shallow water waves in the fully-nonlinear regime. The fully-discrete scheme utilizes the Galerkin / finite-element method based on smooth periodic splines in space, and an explicit fourth-order Runge-Kutta method in time. Computations compared with exact solitary and cnoidal wave solutions show that the scheme achieves the optimal orders of accuracy in space and time. These computations also show that the stability of this scheme does not impose very restrictive conditions on  the temporal stepsize. In addition, solitary, cnoidal, and dispersive shock waves are studied in detail using this numerical scheme for the Serre system and compared with the  `classical' Boussinesq system for small-amplitude shallow water waves. The results show that the interaction of solitary waves in the Serre system is more inelastic. The efficacy of the numerical scheme for modeling dispersive shocks is shown by comparison with asymptotic results. These results have application to the modeling of shallow water waves of intermediate or large amplitude.

\bigskip
\noindent \textbf{\keywordsname:} Green--Naghdi equations; traveling waves; undular bores

\end{abstract}


\maketitle
\tableofcontents
\thispagestyle{empty}

\section{Introduction}

\begin{figure}
\centering
{\includegraphics[width=0.79\textwidth]{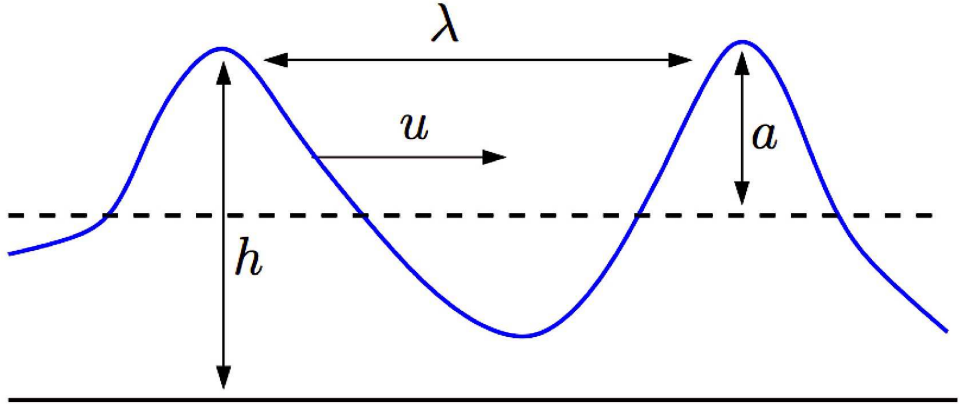}}
\caption{\small\em Sketch of surface water waves for the and variables in the Serre~\eqref{eq:Serre} and cB~\eqref{eq:cB} systems. $u$ is depth-averaged horizontal velocity, $\eps \eta$ is wave height above an undisturbed level (dashes), $h$ is bottom depth, and $\lambda$ a typical wavelength.}
\label{fig:waterwave}
\end{figure}

The propagation of waves on the free surface of an ideal irrotational f{}luid under the force of gravity are governed by Euler's equations \cite{Whitham1999}. Solving Euler's equations is very dif{}ficult, because of the free surface. There is a hierarchy of asymptotic approximations of Euler's equations that do not depend on a free surface. In particular, the propagation of waves in shallow water is governed by the Serre equations, also known as the Su--Gardner equations or Green--Naghdi equations (cf. \cite{Serre1953, Serre1953a, Su1969, Green1976, LordRayleigh1876}), which we shall refer to as the Serre system. In dimensionless variables it reads
\begin{subequations}\label{eq:Serre}
\begin{eqnarray}\label{eq:Serre1}  
\eta_t + u_x + \eps (\eta u)_x &=& 0, \\*[2mm]
\label{eq:Serre2} 
u_t+\eta_x+\eps uu_x-\frac{\sigma^2}{3h} \left[h^3(u_{xt}+\eps uu_{xx}-\eps (u_x)^2)\right]_x &=& 0~,
\end{eqnarray}
\end{subequations}
where 
\begin{equation}\label{eq:h} 
h(x,t) \doteq 1+\eps \eta~.
\end{equation}
Here $x$ is the spatial variable, $t$ is time, $u(x,t)$ is the depth-averaged horizontal velocity of the fluid, $\eps \eta(x,t)$ is the wave height above an undisturbed level of zero elevation, $h(x,t)$ is the total depth of the f{}luid with respect to a horizontal bottom (at a normalized elevation of $-1$ from the undisturbed water level), $\sigma$ is the ratio between the  typical depth $d$ and  wavelength $\lambda$, and $\eps$ is the ratio between the typical amplitude $a$ and bottom depth $d$ (i.e. $\sigma=d/\lambda$ and $\eps=a/d$). See sketch in Fig.~\ref{fig:waterwave}.

The Serre system can be derived as an asymptotic approximation of the Euler's equations under the assumption of shallow water, or long wavelength or weakly dispersive  regime, \ie, $\sigma\ll 1$. Importantly, System~\eqref{eq:Serre} does not assume small amplitude waves, \ie, $\eps$ can be large. For \emph{small-amplitude} or weakly nonlinear shallow water waves, \ie, when $\eps\ll1$, $\sigma\ll 1$, and $\eps/\sigma^2 = \O(1)$, the Serre system reduces to the \acf{cB} system (cf.~\cite{BCS} for related equations), 
\begin{subequations} \label{eq:cB}
\begin{eqnarray}\label{eq:cB1} 
\eta_t + u_x+\eps (\eta u)_x &=& 0~, \\*[2mm]
\label{eq:cB2} 
u_t+\eta_x+\eps uu_x-\frac{\sigma^2}{3}u_{xxt} &=& 0~.
\end{eqnarray}
\end{subequations}

Comparing the cB and~Serre systems, Eqs.~\eqref{eq:cB1} and~\eqref{eq:Serre1} are the  same. This is known as the mass conservation equation. However, Eq.~\eqref{eq:Serre2}  contains higher order nonlinear-dispersive terms compared with~\eqref{eq:cB2}. For this reason, the Serre system is often called {\bf fully-nonlinear} shallow-water equations. Though the two systems are similar, the small amplitude assumption underlying the~cB system appears to be too restrictive for model waves of large, or even intermediate, amplitude. Physically, when water waves approach regions of small depth, it is common that their amplitude increases. Therefore, the Serre system is potentially more appropriate for the approximation of long waves in shallow water and also in the nearshore zone.

There are many studies of the cB system and related Boussinesq-type systems. However, there are much fewer studies of the Serre system, in part because it is considerably harder to solve numerically. As a result, many properties of the solutions of the Serre system are unknown or remain unclear. In particular, few numerical schemes have been developed for the Serre system, based on either the finite dif{}ference (cf.~\cite{Seabra-Santos1987}), hybrid finite-dif{}ference finite-volume  (cf.~\cite{ChazelLannes2010, Bonneton2011}), and pseudospectral (cf.~\cite{Dutykh2011a}) methods. All of these methods can be {\bf formally} very accurate. However, they suf{}fer from aliasing or spurious dissipative ef{}fects, due to the approximation of the nonlinear terms. This is usually not a major handicap when solving the cB system, but it becomes debilitating when solving the Serre system for large amplitude and / or rapidly oscillating waves, as in such cases the actual error is large, even when using a fine grid. Even worse, such schemes often fail to converge.

In this study, we design and implement a computational scheme based on the standard \acf{FEM}. One of the main advantages of this method is that it is {\bf non-dissipative}. Another advantage is the sparsity of the resulting linear systems. For this reason, as we show, the \acs{FEM} scheme for~\eqref{eq:Serre} achieves the optimal (formal) order of accuracy even for large amplitude solitary, cnoidal and dispersive shock waves. To achieve  this, we use cubic splines for the semi-discretization in space and the classical fourth order \acf{RK} method for time integration. A similar scheme was studied for a variety of Boussinesq-like systems, cf.~\cite{DMII}, and appears to be highly accurate and ef{}ficient in a measure that makes the method to be ideal for the study of the dynamics of solitary waves. However, for the Serre system, the dependence of the dispersive terms in~\eqref{eq:Serre2} on the unknown function $h(x,t)$ makes the numerical integration considerably more dif{}ficult compared with Boussinesq-like systems. In particular, for time integration, a {\bf mass matrix} needs to be assembled at each time step. At every intermediate \acs{RK} step, two linear systems based on the mass matrix need to be solved. This is a costly computation, yet, in spite of this drawback, the high accuracy of this scheme makes it a strong candidate for computational modeling of the Serre system.

The paper is organized as follows. Section~\ref{sec:analytic} recaps some of the analytical results for solitary, cnoidal, and dispersive shock wave solutions of the Serre system, which serve to validate the computational scheme. Section~\ref{sec:scheme} presents the fully-discrete schemes for the Serre and cB systems. Section~\ref{sec:valid} validates the convergence, accuracy, and numerical stability of the method. Section~\ref{sec:experiments} presents  computational studies of interacting solitary waves and dispersive shock waves.

\subsection{Remarks}

System~\eqref{eq:cB} was originally derived (in a slightly dif{}ferent form) by Boussinesq~\cite{Boussinesq1872} and it is a special case of a class of Boussinesq-type models derived by Bona, Chen and Saut, \cite{BCS}. This system is often used for studying two-way propagation of small amplitude, long waves~\cite{Peregrine1967}.

The Serre system~\eqref{eq:Serre} was originally derived from Euler's equations in one spatial dimension by Serre~\cite{Serre1953, Serre1953a}. It was re-derived several times later, including independently by Su and Gardner~\cite{Su1969}. See also the review by Barth\'elemy~\cite{Barthelemy2004}~. Green and Naghdi~\cite{Green1976} generalized this system to two spatial dimensions with an uneven bottom, see also~\cite{Seabra-Santos1987, Wei1995}. Recently, Lannes and Bonneton~\cite{Lannes2009} derived and justified several asymptotic models of surface waves including~\eqref{eq:Serre}. It is worth mentioning that other systems of a similar ilk have been derived, for example, with improved dispersion characteristics~\cite{Lannes2009, Zhang2013} and with surface tension~\cite{Dias2010}. In principe, all these models can be discretized by numerical methods similar to the methods presented bellow. However, further studies will be required to test the efficacy of the ensuing schemes.

\section{Analytical properties of the Serre system}\label{sec:analytic}

Below we recap several analytical properties of System~\eqref{eq:Serre}, which serve as benchmarks for our computational scheme.

\subsection{Special solutions of the Serre system}

System~\eqref{eq:Serre} admits solitary and cnoidal wave solutions in closed form (cf.~\cite{Serre1953, El2006, Carter2011}). Below we recap these solutions for arbitrary $\eps$ and $\sigma$. It is convenient to express the solutions of \eqref{eq:Serre} in terms of $(h, u)$ rather than $(\eta, u)$. The two ways are equivalent in light of \eqref{eq:h}.

The two-parameter family of solitary wave solutions of~\eqref{eq:Serre} that travel with a constant speed $c_s$ can then be written in the moving frame $\xi=x-c_s t$ as
\begin{subequations}\label{eq:soliton}
\begin{eqnarray}\label{eq:h_sol}  
h_{sol}(\xi) &=& \frac{1}{\sigma} \left[ a_0 + a_1{\sech}^2(K_s\, \xi) \right] \\*[2mm]
\label{eq:u_sol} 
u_{sol}(\xi) &=& \frac{c_s}{\eps}\left[ 1 - \frac{a_0}{\sigma h_{sol}(\xi)} \right]
\end{eqnarray}
\end{subequations}
where 
$$
K_s=\sqrt{\frac{3 a_1}{4\sigma a_0^2 c_s^2}}~, \qquad c_s=\sqrt{\frac{a_0+a_1}{\sigma}}~,
$$
and $a_0,a_1$ are positive (but otherwise arbitrary) constants. Choosing $a_0=\sigma$ gives the solitary waves that decay to the background average water depth. For the simulations, we choose $\eps = a_0 = \sigma = 1$ and various values of the speed $c_s$. Then $a_1$ and $K_s$ can be determined from the above formulae.

Similarly, the three-parameter cnoidal wave solutions of~\eqref{eq:Serre} can be written as
\begin{subequations}\label{eq:cn}
\begin{eqnarray}\label{eq:h_c}  
h_c(\xi) &=&  \frac{1}{\sigma} \left[ a_0+a_1 {\rm dn}^2(K_c\,\xi, k)\right], \\*[2mm]
\label{eq:u_c} 
u_c(\xi) &=& \frac{c_s}{\eps}\left[ 1-\frac{h_0}{\sigma h_c(\xi)} \right]
\end{eqnarray}
\end{subequations}
where ${\rm dn}$ denotes the Jacobi elliptic function and
\begin{eqnarray*}
h_0 &=& a_0+a_1\frac{E(m)}{K(m)}~,\\
K_c &=& \frac{\sqrt{3a_1}}{2\sqrt{a_0(a_0+a_1)(a_0+(1-k^2)a_1)}}~,\\*[2mm]
c_s &=& \sqrt{\frac{a_0(a_0+a_1)(a_0+(1-k^2)a_1)}{\sigma h_0^2}}~.
\end{eqnarray*}
$K(m)$ and $E(m)$ denote the complete elliptic integrals of the first and second kind, respectively, $k \in [0, 1]$, $m = k^2$, and $a_0$, $a_1$ are positive constants. Here, $a_0$, $a_1$ and $m$ (or $k$) are arbitrary.

It is remarkable that such closed form solutions of the Serre have been found and even more remarkable that such closed form solutions \emph{have not} been found for the cB system~\eqref{eq:cB}. Nevertheless, it has been proven that the cB system admits solitary and cnoidal wave solutions (cf.~\cite{Chen2000, Chen2007}).

\subsection{Dispersive shock waves}\label{sec:DSW}

\begin{figure} 
\centering
{\includegraphics[width=0.86\textwidth]{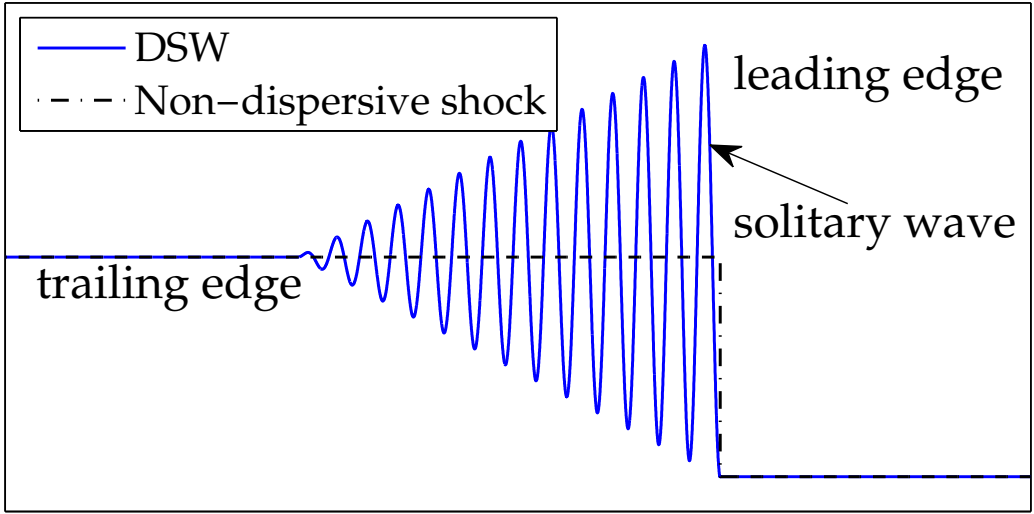}}
\caption{\small\em Sketch of dispersive shock wave (DSW, solid) and non-dispersive shock (dashes).}
\label{fig:DSW}
\end{figure}

When the dispersive terms in the Serre or cB systems are neglected, the resulting non-dispersive equations can give rise to supersonic shock waves, \ie, a discontinuous solution. When such shocks are regularized by dissipative ef{}fects, this gives rise to classical or viscous shocks, which are characterized by a rapid and monotonic change in the f{}low properties. On the other hand, in systems where dissipation is negligible compared with dispersion, the dispersive ef{}fects give rise to \acf{DSWs}. DSWs are characterized by an expanding train of rapidly-oscillating waves (see sketch in Fig.~\ref{fig:DSW}). The leading edge of a DSW possesses large amplitude waves, which decay to linear waves at the trailing edge. DSWs have been studied for several decades, originally in the context of the KdV equation (cf.~\cite{Benjamin1954, Whitham1967, Gurevich1974} for some of the early works). In particular, these studies show that, using Whitham's averaging method, the largest-amplitude oscillation in the leading edge is well-approximated with a solitary wave.

Recently, DSWs were studied analytically and computationally in fully-nonlinear dispersive shallow water systems (cf.~\cite{El2006, El2008, LeMetayer2010, Esler2011}). In particular, the asymptotic dynamics of the leading edge solitary wave of a simple DSW were studied  in~\cite{El2006, El2008}. Below we recap some of those results. We use these results to test the non-dissipativity of the numerical schemes.

Consider the Serre system~\eqref{eq:Serre} with $\eps = \sigma = 1$. As above, writing the solution of the Serre system in terms of $h$ and $u$, a {\bf simple DSW} traveling to the right is generated using the Riemann initial data
\begin{equation}\label{eq:DSW-init}
h(x,0)=\left\{
\begin{array}{ll}
  h^{-}, & \mbox{ for } x < 0\\
  h^{+}, & \mbox{ for } x > 0
\end{array}
\right. ,
\quad 
u(x,0)=\left\{
\begin{array}{ll}
  u^{-}, & \mbox{ for } x < 0\\
  u^{+}, & \mbox{ for } x > 0
\end{array}
\right.
\end{equation}
with the compatibility condition (Riemann invariant)
\begin{equation}\label{eq:u-}
  \frac{u^{-}}{2} - \sqrt{h^{-}} = \frac{u^{+}}{2} - \sqrt{h^{+}}.
\end{equation}
Following~\cite{El2006, El2008}, we assume that the initial jump or total depth variation is small, \ie,
\begin{equation}\label{eq:delta}
  0 < \delta \ \doteq \ \frac{h^{-}}{h^{+}} -1 \ \ll\  1~.
\end{equation}
Then, to leading order in $\delta$ and for a large propagation time, the leading-edge of the DSW is well approximated with a solitary wave given by~\eqref{eq:soliton} with $a_0 = 1$ and amplitude (relative to the constant elevation) and speed
\begin{subequations}\label{eq:DSW-simple}
\begin{eqnarray}\label{eq:DSW-simple-a1}
  a_1^{\rm simple} &\sim& 2\delta + \frac{1}{6}{\delta}^2~,\\*[2mm]
\label{eq:DSW-simple-cs}
  c_s^{\rm simple} &\sim& 1+\delta-\frac{5}{12}{\delta}^2~, 
\end{eqnarray}  
\end{subequations}
respectively.

We also consider the {\bf dam-break problem}. In this case, the initial data for $h(x,0)$ is the same~\eqref{eq:DSW-init}. However, there is no f{}low at $t = 0$, \ie, $u(x,0) = 0$. As shown in~\cite{El2006}, this generates two counter-propagating DSWs, one on each side of the ``dam'', and two rarefaction waves that travel towards the center. Similarly to a simple DSW, the asymptotic amplitude and speed for the leading-edge solitary wave in each DSW is
\begin{subequations}\label{eq:DSW-dam}
\begin{eqnarray}\label{eq:DSW-dam-a1}
a_1^{\rm dam} &\sim& \delta - \frac{1}{12}{\delta}^2~,\\*[2mm]
\label{eq:DSW-dam-cs}
c_s^{\rm dam} &\sim& 1+ \frac{\delta}{2} -\frac{1}{6}{\delta}^2~, 
\end{eqnarray}  
\end{subequations}
respectively.

\subsection{Hamiltonian conservation}

A fundamental property of the Serre system is its Hamiltonian structure, cf.~\cite{Johnson2002, Li2002}. For any solution $(h,u)$, the energy functional (or Hamiltonian)
\begin{equation}\label{eq:H} 
  \H(t) \equiv \H[h,u] = \int_{-\infty}^{\infty} \left[
    \eps h u^2+\frac{\eps\sigma^3}{3}h^3u_x^2+\frac{1}{\eps}(h-1)^2 \right]\ dx, 
\end{equation}
is conserved in the sense that $\H(t) = \H(0)$ for all $t > 0$ and up to the maximal time $T$ of the existence of the solution. In contrast, the cB system~\eqref{eq:cB} does not possess a Hamiltonian structure and its solutions do not conserve an energy functional~\cite{BCS}.

\section{The \acs{FEM} scheme}\label{sec:scheme}

In this section we present a \acs{FEM} for the initial boundary value problem (IBVP) comprised of System~\eqref{eq:Serre} subject to periodic boundary conditions. Here and in the computations we choose $\eps = \sigma = 1$. We make this choice in order to simulate large amplitude waves, so as to ``push'' the scheme to its limit. For this reason, the $\eps$ and $\sigma$ are dropped from the equations below. It is also convenient to rewrite~\eqref{eq:Serre1} in terms of $(h, u)$ rather than $(\eta, u)$. This is done using~\eqref{eq:h} and yields the IBVP
\begin{subequations}\label{eq:IBVP} 
\begin{eqnarray}\label{eq:IBVP1}
h_t + (h u)_x &=&0, \\*[2mm]
\label{eq:IBVP2} 
u_t+ h_x + uu_x-\frac{1}{3h} \left[h^3(u_{xt} +  uu_{xx} - (u_x)^2)\right]_x &=& 0,
\end{eqnarray}
\begin{eqnarray}\label{eq:bc1}
\partial_x^i h(a,t) &=& \partial_x^i h(b,t),   \quad i=0,1,2,\ldots, \\*[2mm]
\label{eq:bc2}
\partial_x^i u(a,t) &=& \partial_x^i u(b,t),   \quad i=0,1,2,\ldots, \\*[2mm]
\label{eq:ic}
h(x,0) &=& h_0(x), \\*[2mm] 
\label{eq:ic2}
u(x,0) &=& u_0(x),
\end{eqnarray}
\end{subequations}
where $x \in (a, b) \subset \R$ and $t \in [0,T)$. We shall assume that~\eqref{eq:IBVP} possesses a unique solution, such that $h$ and $u$ are suf{}ficiently smooth and, for any $t\in[0,T]$, in a suitable Sobolev space with periodic boundary conditions, \ie,
$$
h(x,\cdot) \in H^s, \quad u(x,\cdot) \in H^{s+1}, \quad H^s \equiv  H^s_{per}(a,b)~,
$$
where $s\ge 2$ (see~\cite{Israwi2011} for sharper results). Here and below, $\|\cdot\|_s$ denotes the standard norm in $H^s$. We also use the inner product in $L^2 \equiv H^0$, denoted by $(\cdot,\cdot)$.

\subsection{Spatial discretization}

We denote the spatial grid by $x_i=a+i\, \Delta x$, where $i=0,1,\cdots, N$, $\Delta x$ is grid size, and $N\in\N$, such that $\Delta x = (b-a)/N$. Let $(\h,\u)$ be the corresponding spatially discretized solution. The \acf{FEM} seeks a weak solution of~\eqref{eq:IBVP}, \ie, $\h(x,t)$ and $\u(x,t)$ in $C^1(0,T;S)$, for a suitable finite-dimensional space $S$. We shall consider the space of the smooth periodic splines
$$
S^r = \{\phi\in \left. C^{r-2}_{per}[a,b] \right| \phi|_{[x_i,x_{i+1}]}\in\mathbb{P}^{r-1}, \quad 0\leq i\leq N-1\},
$$
where  $r\in\N$,
$$
C_{per}^{r} \doteq \{ \left. f\in C^r[a,b] \right|f^{(k)}(a)= f^{(k)}(b), \quad 0\leq k\leq r\},
$$
and $\mathbb{P}^{k}$ are the polynomials of degree at most $k$. In particular, we shall use cubic splines, which correspond to $S^4$, \ie, $r = 4$. Henceforth, we shall denote $S \equiv S^4$. Note that the periodic boundary conditions~\eqref{eq:bc1}--\eqref{eq:bc2} are satisfied automatically by this choice.

To state the associated weak problem, let $\phi\in S$ be an arbitrary test function. It turns out to be convenient to multiply~\eqref{eq:IBVP2} by $h$ and group together the first term in this equation with the first term in the brackets. Using integration by parts several times, gives the {\bf semi-discrete problem}
\begin{subequations}\label{eq:weak} 
\begin{eqnarray}\label{eq:weak1}
(\h_t,\phi) +  ((\h \u)_x,\phi) &=& 0, \\*[2mm]
\label{eq:weak2} 
\BS(\u_t,\phi;\h) + (\h( \h_x + \u\u_x),\phi) + \frac{1}{3} \left(\h^3(\u\u_{xx} - (\u_x)^2),\phi_x\right) &=& 0~, \\[2mm]
\end{eqnarray} 
where the bilinear form is defined for a fixed $\h$ (and substituting $\psi\equiv \u_t$) as
\begin{equation}\label{eq:B}
\BS(\psi,\phi;\h) \ \doteq \ (\h \psi,\phi) + \frac{1}{3}(\h^3 \psi_x,\phi_x)~, 
\end{equation}
and the initial conditions are
\begin{equation}\label{eq:weak-init} 
\h(x,0) \ =\ \P \{ h_0(x)\}~, \qquad \u(x,0)=\P \{u_0(x)\}
\end{equation}
\end{subequations}
where $\P$ is the $L^2$ projection onto the $S$ defined by $(\P v,\phi)=(v,\phi)$, for all $\phi\in S$.

The bilinear form~\eqref{eq:B} plays a key role in the \acs{FEM}. Assuming that $h$ is bounded as $h_0 \le h(x,t)\le \infty$ for some constant $h_0>0$ (the so called ``wet bottom'' assumption), this bilinear form is bounded and coercive. Specifically, for all $\phi,\psi\in H^1$, there exist $c_1,c_2$, such that
\begin{equation}\label{eq:bounded}
 \BS(\psi,\phi;\h)\leq c_1\|\phi\|_1\|\psi\|_1~, \qquad \BS(\phi,\phi;\h)\geq c_2\|\phi\|_1^2~.
\end{equation}
These two properties are of fundamental importance for the \acs{FEM} scheme. In particular, it follows from~\eqref{eq:bounded} that the corresponding linear systems are not singular.

\subsection{Temporal discretization}

Upon choosing appropriate basis functions for $S$, System~\eqref{eq:weak}  represents a system of ordinary dif{}ferential equations (ODEs). For time integration, we employ the classical, explicit, four-stage, fourth-order \acf{RK} method, which is described by the following \emph{tableau}:
\begin{equation}\label{tab:RK}
\begin{tabular}{c | c}
  $A$ & $b$ \\ \hline
  $\tau$ &  
\end{tabular} \ = \
\begin{tabular}{c c c c| c}
  0 & 0 & 0 & 0 & 1/6\\
  1/2 & 0 & 0 & 0 & 1/3 \\ 
  0 & 1/2 & 0 & 0 & 1/3 \\
  0 & 0 & 1/2 & 0 &1/6\\\hline
  0 & 1/2 & 1/2 & 1 &
\end{tabular}~.
\end{equation}
We use a uniform time-step $\Delta t$, such that $\Delta t=T/K$ for a suitable $K\in\N$. The temporal grid is then $t^n=n\, \Delta t$, where $n=0,1,\cdots, K$. Given the ODE $y' = \Phi(t,y)$, one step of this fourth-stage \acs{RK} scheme (with $y^n$ approximating $y(t^n)$) is

\begin{algorithmic}
  \For{$i = 1 \to 4$} 
   \State $\tilde{y}^i ~~~=\ y^n+\sum_{j=1}^{i-1}a_{ij}\,y^{n,j}$ \vspace{1mm}
   \State $y^{n,i}  \ =\ \Phi(t^{n,i},\tilde{y}^i),  \quad \mbox{evaluated at~}  t^{n,i} \equiv t^n+\tau_i \Delta t $
  \EndFor 
  \State $y^{n+1}=y^n+\Delta t\sum_{j=1}^4 b_j\,y^{n,j}$~,  \vspace{2mm}
\end{algorithmic}
where $a_{ij}$, $\tau_i$, $b_i$ are given in Table~\ref{tab:RK}. Applying this scheme to~\eqref{eq:weak} and denoting by $H^n$ and $U^n$ the fully discrete approximation in $S$ of $h(\cdot,t^n)$, $u(\cdot,t^n)$, respectively, leads to Algorithm~\ref{alg:Serre}.

\begin{algorithm}
\caption{\small\em Time-marching \acs{FEM} scheme for Serre IBVP~\eqref{eq:weak}.}
\begin{algorithmic}\label{alg:Serre}
  \State $H^0 \ =\  \P \{h_0\}$ \vspace{2mm}
  \State $U^0 ~\, =\  \P \{u_0\}$ \vspace{2mm}
  \For{$n = 0 \to N-1$}  \vspace{2mm}
  \For{$i = 1 \to 4$}  \vspace{2mm}
    \State $\tilde{H}^i \ =\ H^n \ +\ \sum_{j=1}^{i-1}a_{ij}\,H^{n,j}$\vspace{2mm}
    \State $\tilde{U}^i ~\, =\ U^n \ + \ \sum_{j=1}^{i-1}a_{ij}\,U^{n,j}$\vspace{2mm}
    \State $(H^{n,i},\phi) \hspace{5.5mm} \ =\ -((\tilde{H}^{i}\tilde{U}^{i})_x,\phi), 
    \quad\mbox{evaluated at~} t^{n,i} \equiv t^n+\tau_i \Delta t $\vspace{2mm}
    \State $\BS(U^{n,i},\phi;\tilde{H}^{i}) \ =\ -((\tilde{H}^{i}(\tilde{H}_x^{i}+\tilde{U}^{i}\tilde{U}^{i}_x),\phi) \ -\ \frac{1}{3}(\tilde{H}^{i})^3(\tilde{U}^{i}\tilde{U}_{xx}^{i}-(\tilde{U}_x^{i})^2,\phi_x)$\vspace{2mm}
  \EndFor \vspace{2mm}
  \State $H^{n+1} \ =\ H^n \ + \ \Delta t\sum_{j=1}^4 b_j\,H^{n,j}$ \vspace{2mm}
  \State $U^{n+1} ~\, =\ U^n \ +  \ \Delta t\sum_{j=1}^4 b_j\,U^{n,j}~$\vspace{2mm}
  \EndFor 
\end{algorithmic}
\end{algorithm}

Given a basis $\{\varphi_i\}$ of $S$, the implementation of Algorithm~\ref{alg:Serre}  requires solving at each time step the following linear systems.
\begin{enumerate}
  \item Four linear systems with the {\bf time-independent} matrix $(\varphi_i,\varphi_j)$;
  \item Four linear systems with the {\bf time-dependent} matrix $\BS(\varphi_i,\varphi_j;h)$~.
\end{enumerate}
All these matrices are cyclic and symmetric due to the periodic boundary conditions. They consist of a seven-diagonal band and two $3\times 3$ triangular blocks on the upper right and lower left corners. To solve these systems, we use the direct method described in \cite{BDKMc}, which is analogous to the Sherman-Morrison-Woodbury method. To approximate the inner products, we use the Gauss-Legendre quadrature with 5 nodes per $\Delta x$.

We note that the above algorithms are almost identical in the case of other boundary conditions but the convergence and the stability properties will be different. In the analogous case of the \acs{cB} system the convergence of the \acs{FEM} has optimal rates of convergence in the periodic case, \cite{Antonopoulos2010}, contrary to the suboptimal rates characterized the problem subject to non-periodic boundary conditions, cf. \cite{AD3,AD2013}. For more information on implementation of the Galerkin / Finite Element methods with other boundary conditions see~\cite{Schultz1973}.

\subsection{\acs{FEM} scheme for the cB system}

Below we brief{}ly outline the corresponding scheme for the cB system~\eqref{eq:cB}. See~\cite{AD3, Antonopoulos2010, ADM2} for details. Making the transformation $h(x,t) \mapsto 1+ \eta(x,t)$, the semi-discrete problem for the cB system~\eqref{eq:cB} is
\begin{subequations}\label{eq:cB-semi}
\begin{eqnarray}
(\h_t,\phi)+( (\h\u)_x,\phi) &=& 0 \\*[2mm]
\BB({\u}_{xt},\phi_x)+(\h_x,\phi)+(\u {\u}_x,\phi) &=& 0,
\end{eqnarray}
where, in this case, the bilinear form is defined as
\begin{equation}
 \BB(\psi,\phi) \ \doteq \ (\psi,\phi)+\frac{1}{3}(\psi_x,\phi_x)~.
\end{equation}
\end{subequations}

Using the notation as in Algorithm~\ref{alg:Serre} and denoting the fully-discrete solution by $h(x,t)\to H^n(x)$ and $u(x,t)\to U^n(x)$, the corresponding full-discrete algorithm based on the same \acs{RK} scheme is given by Algorithm \ref{alg:cB}.

\begin{algorithm}
\caption{Time-marching \acs{FEM} scheme for the IBVP of the cB system~\eqref{eq:cB}}
\label{alg:cB}
\begin{algorithmic}[h]
  \State $H^0 \ =\  \P \{h_0\}$ \vspace{2mm}
  \State $U^0 ~\, =\  \P \{u_0\}$ \vspace{2mm}
  \For{$n = 0 \to N-1$}  \vspace{2mm}
  \For{$i = 1 \to 4$}  \vspace{2mm}
    \State $\tilde{H}^i \ =\ H^n \ +\ \sum_{j=1}^{i-1}a_{ij}\,H^{n,j}$\vspace{2mm}
    \State $\tilde{U}^i ~\, =\ U^n \ + \ \sum_{j=1}^{i-1}a_{ij}\,U^{n,j}$\vspace{2mm}
    \State $(H^{n,i},\phi) \hspace{5.5mm} \ =\ -((\tilde{H}^{i}\tilde{U}^{i})_x,\phi), 
    \quad\mbox{evaluated at~} t^{n,i} \equiv t^n+\tau_i \Delta t $\vspace{2mm}
    \State $\BB(U^{n,i},\phi) \ =\ -(\tilde{H}^i_x,\phi)-(\tilde{U}^i\tilde{U}^i_x,\phi)$ \vspace{2mm}
  \EndFor \vspace{2mm}
  \State $H^{n+1} \ =\ H^n \ + \ \Delta t\sum_{j=1}^4 b_j\,H^{n,j}$ \vspace{2mm}
  \State $U^{n+1} ~\, =\ U^n \ +  \ \Delta t\sum_{j=1}^4 b_j\,U^{n,j}~$\vspace{2mm}
  \EndFor 
\end{algorithmic}
\end{algorithm}

As for the \acs{FEM} scheme for the Serre problem, we employ cubic splines and 
the inner products are approximated using a 5-node Gauss--Legendre quadrature. The resulting linear systems are similar with those of Algorithm~\ref{alg:Serre} and are solved using the same numerical method. The key dif{}ference from Algorithm~\ref{alg:Serre} is that {\bf all the matrices in Algorithm~\ref{alg:cB} are time-independent}. Therefore, the matrices are assembled and factorized once and for all at $t=0$.

\subsection{Theoretical considerations}
\label{sec:stab}

For the semi-discrete problem (\ref{eq:cB-semi}), it was proven in~\cite{Antonopoulos2010} that, for appropriate initial conditions and for small values of $\Delta x$, there is a unique semi-discrete solution, which satisfies the estimate
\begin{equation}
  \label{EA6}
  \max_{0\leq t\leq T}(\|h-\tilde{h}\|+\|u-\u\|)\leq C\Delta x^r,
\end{equation}
where the constant $C$ is independent of $\Delta x$. This result also shows that the numerical solution is stable. Furthermore, the same result is valid for the linear \acs{cB} system. Since the linearization of the \acs{cB} and Serre systems is the same, it is implied that the semi-discrete solution of the linearized Serre system is stable. No stability or convergence results are known for the nonlinear (semi- or fully-) discrete schemes for the Serre equations.

\section{Validation of the \acs{FEM} scheme for the Serre system}\label{sec:valid}

In this section we study the spatial and temporal accuracy and ef{}ficiency of the \acs{FEM} scheme for the Serre system, which is presented in Algorithm~\ref{alg:Serre}. To do so, we use various metrics of the solitary and cnoidal wave solutions and Hamiltonian conservation (see Section~\ref{sec:analytic}).

\subsection{Spatial accuracy}\label{sec:spatial}

To test the spatial accuracy of the scheme, we use the exact solitary wave solution~\eqref{eq:soliton} of the Serre system with $\eps = a_0 = \sigma = 1$ and traveling with speed $c_s = 1.5$. The spatial domain is chosen as $x\in [-150, 150]$. This large interval ensures that the solution is practically zero near the endpoints of the interval. To ensure that the errors incurred by the temporal integration are negligible, we take $\Delta t/\Delta x = 0.1\ $.

Tables~\ref{T1}--\ref{T3} show the normalized errors of the computed solutions evaluated at $T = 100$. These errors are defined as
\begin{equation}\label{eq:error}
  \err_s[F] \ \doteq \ \frac{ \| F(x,T;\Delta x) -\Fex(x,T) \|_s }{ \|\Fex(x,T)\|_s }~,
\end{equation}
where $F = F(\cdot;\Delta x)$ is the computed solution, \ie, either $H\approx h(x,T)$ or $U\approx u(x,T)$, $\Fex$ is the corresponding exact solitary wave solution with the same parameters (see~\eqref{eq:h_sol}--\eqref{eq:u_sol}), and $s = 0, 1, \infty$ corresponds to the $L^2$, $H^1$, and $L^\infty$ norms, respectively.

For the calculations of the $L^\infty$ norm and related variables mentioned in the following sections, we recover {\bf location of the peak amplitude} curve of the solution $h(x,t)$. This curve, denoted by $\xa(t)$, is defined via
\begin{equation}\label{eq:x_peak}
  \left. \frac{d}{dx}h(x,t)\right|_{x=\xa(t)} = 0~.
\end{equation}
To compute $\xa(t)$, we use Newton's method. As an initial guess, we use the quadrature node at which $H^n(x)$ attains a maximum over all the quadrature nodes. This ensures that $H^n(\xa)$ is the global maximum. Usually, only a few iterations are needed to achieve $\xa$ with a tolerance of $10^{-13}$.

Tables~\ref{T1}--\ref{T3} also show the corresponding calculated rates of convergence, defined as 
$$
\mbox{rate for~}  \err_s[F]  \ \doteq \ 
\frac{ \ln\left( \err_s[F(\cdot;\Delta x_{k-1})] / \err_s[F(\cdot;\Delta x_k)] \right) }  
{ \ln\left(\Delta x_{k-1}/\Delta x_{k}\right)}~,
$$
where $\Delta x_{k}$ is the grid size listed in row $k$ in each table.

These tables show that the rates using the $L^2$ and $L^\infty$ norms approach $4$, whereas, the rates using the $H^1$ norm approaches $3$. These results indicate that the \acs{FEM} scheme achieves the optimal orders of convergence. Moreover, one might expect the emergence of large errors (in space and/or time) due to the following challenging conditions:
\begin{enumerate}
\item The high-order nonlinear dispersive terms in~\eqref{eq:Serre}.
\item The strongly nonlinear and dispersive regime ($\eps = \sigma = 1$).
\item The use of an explicit \acs{RK} method.
\end{enumerate}
Yet, in spite of these challenging conditions, Tables~\ref{T1}--\ref{T3} show that the actual errors are very small, even when using relatively large grid sizes. {\bf Hence, these results show that this scheme is highly ef{}ficient.}

\begin{table}
\caption{\small\em Spatial errors and  rates of convergence for a solitary wave solution using the $L^2$ norm [Eq.~\eqref{eq:error}  with $s=2$].}\label{T1}
\begin{tabular}{cccccc}
\hline
$\Delta x$ & $\Delta t$ & $\err_2[H]$ & rate for $\err_2[H]$ & $\err_2[U]$ & rate for $\err_2[U]$\\
\hline
$ 0.50$ & $0.05$ &  $0.1970\times 10^{-4}$ & -- & $0.5669\times 10^{-4}$ & -- \\
$0.25$ & $0.25$ & $0.7989\times 10^{-6}$ &  $4.6240$ & $0.2153\times 10^{-5}$ & $4.7183$\\
$0.10$ & $0.01$ & $0.1798\times 10^{-7}$ &  $4.1402$ & $0.4973\times 10^{-7}$ & $4.1124$\\
$0.08$ & $0.008$ & $0.7298\times 10^{-8}$ &  $4.0420$ & $0.2018\times 10^{-7}$ & $4.0403$\\
$0.05$ & $0.005$ & $0.1102\times 10^{-8}$ &  $4.0224$ & $0.3043\times 10^{-8}$ & $4.0257$\\
\hline
\end{tabular}
\end{table}

\begin{table}
\caption{\small\em Same as Table~\ref{T1} using the $H^1$ norm.}\label{T2}
\begin{tabular}{cccccc}
\hline
$\Delta x$ & $\Delta t$ & $\err_2[H]$ & rate for $\err_2[H]$ & $\err_2[U]$ & rate for $\err_2[U]$\\
\hline
$ 0.50$ & $0.05$ &  $0.1951\times 10^{-3}$ & -- & $0.41324\times 10^{-3}$ & -- \\
$0.25$ & $0.25$ & $0.1873\times 10^{-4}$ &  $3.3805$ & $0.4261\times 10^{-4}$ & $3.2775$\\
$0.10$ & $0.01$ & $0.1111\times 10^{-5}$ &  $3.0830$ & $0.2601\times 10^{-5}$ & $3.0515$\\
$0.08$ & $0.008$ & $0.5660\times 10^{-6}$ &  $3.0236$ & $0.1327\times 10^{-5}$ & $3.0143$\\
$0.05$ & $0.005$ & $0.1374\times 10^{-6}$ &  $3.0122$ & $0.3230\times 10^{-6}$ & $3.0073$\\
\hline
\end{tabular}
\end{table}

\begin{table}
\caption{\small\em Same as Table~\ref{T1} using the $L^\infty$ norm.}\label{T3}
\begin{tabular}{cccccc}
\hline
$\Delta x$ & $\Delta t$ & $\err_\infty[H]$ & rate for $\err_\infty[H]$ & $\err_\infty[U]$ & rate for $\err_\infty[U]$\\
\hline
$ 0.50$ & $0.05$ &  $0.4228\times 10^{-3}$ & -- & $0.5315\times 10^{-4}$ & -- \\
$0.25$ & $0.25$ & $0.2101\times 10^{-4}$ &  $4.3309$ & $0.2882\times 10^{-5}$ & $4.2049$\\
$0.10$ & $0.01$ & $0.4887\times 10^{-6}$ &  $4.1046$ & $0.7123\times 10^{-7}$ & $4.0383$\\
$0.08$ & $0.008$ & $0.1988\times 10^{-6}$ &  $4.0291$ & $0.2893\times 10^{-7}$ & $4.0381$\\
$0.05$ & $0.005$ & $0.3013\times 10^{-7}$ &  $4.0148$ & $0.4373\times 10^{-8}$ & $4.0199$\\
\hline
\end{tabular}
\end{table}

\subsection{Temporal accuracy and stability}\label{sec:temporal}

In order to study the temporal accuracy, we use the same solitary wave solutions as above. Here we take $\Delta t = \Delta x$ for various values of $\Delta x = (b-a)/N$. This choice for $\Delta t$ and $\Delta x$ is sufficient to estimate the temporal order of accuracy for the following reason. Since our spatial discretization is $4^{\rm th}$-order, we may assume that the scheme's total error at some final time $t=T$ scales as
\begin{equation}\label{eq:err1}
  \err_s[F] \doteq \|f-F\| =  C(\Delta x^4 + \Delta t^r)~,
\end{equation}
where $f = f(\cdot, T; \Delta x, \Delta t)$ stands for the computed solution, $F(\cdot,T)$ stands for the exact solution, $C$ is a constant, and $r$ is the temporal convergence rate.  Since our schemes use a $4^{th}$-order Runge-Kutta method, it is expected that $r\le 4$. By choosing $\Delta x=\Delta t \ll 1$ and using~\eqref{eq:err1}, the total error scales as
\begin{equation}\label{eq:err2}
  \err_s[F] = \|f-F\| = C\Delta t^r~.
\end{equation}
Choosing two different values of $\Delta t$, \ie, $\Delta t_{k-1}$ and $\Delta t_k$, gives
\begin{equation}\label{eq:err3}	
  \err_s[F(\cdot;\Delta t_{k-1})] = C\Delta t_{k-1}^r~, \quad \err_s[F(\cdot;\Delta t_{k})] = C\Delta t_{k}^r~.
\end{equation}
Taking the ratio of these two errors and solving for temporal convergence rate, $r$, yields
\begin{equation}\label{eq:temprate}
\mbox{rate for~} \err_s[F] \ \doteq \ 
\frac{\ln\left(\err_s[F(\cdot;\Delta t_{k-1})] / \err_s[F(\cdot;\Delta t_k)] \right)}
{\ln\left(\Delta t_{k-1}/\Delta t_{k}\right)}~.
\end{equation}

Tables~\ref{T4}--\ref{T6} present the errors defined in~\eqref{eq:error} and the corresponding rates of convergence, defined by (\ref{eq:temprate}), where $\Delta t_{k}$ is the grid size listed in row $k$. These tables show that the \acs{FEM} scheme achieves the optimal temporal rate of convergence in all three norms. The actual errors are fairly small as well. Moreover, one might expect the scheme to be conditionally stable, due to the complexity of the problem and the use of an  explicit \acs{RK} method. Yet, the  scheme converges in spite of these challenging conditions and the large temporal grid size ($\Delta t=\Delta x$). We note that it has been proven that a similar \acs{FEM} scheme is \emph{unconditionally stable} for several types of Boussinesq systems~\cite{ADM2}. Although we do not have a proof that this fully-discrete problem is unconditionally stable, these results show that {\bf the stability of this scheme does not impose restrictive conditions on $\Delta t$} but mild conditions of the form $\Delta t< C\ \Delta x$ are adequate for the solutions to remain stable. This property is further explored in Section~\ref{sec:stab}.

\begin{table}
\caption{\small\em Same as Table~\ref{T1} for the temporal errors. $N$ and $M$ are the  number of spatial and temporal grid points, respectively.}\label{T4}
\begin{tabular}{cccccc}
\hline
$N$ & $M$ & $\err_2[H]$ & rate for $\err_2[H]$ & $\err_2[U]$ & rate for $\err_2[U]$\\
\hline
$200$ & $500$ &  $0.1824 \times 10^{-1}$ & -- & $0.9673\times 10^{-2}$ & -- \\
 $400$ & $1000$ &  $0.7114 \times 10^{-3}$ & $4.6808$ &  $0.3759 \times 10^{-3}$ & 4.6853 \\
 $800$ & $2000$ &  $0.3055 \times 10^{-4}$ & $4.5411$ &  $0.1669 \times 10^{-4}$ &  4.4935 \\
$1600$ & $4000$ &  $0.1496\times 10^{-5}$ & $4.3521$  & $0.8296\times 10^{-6}$ &  4.3303 \\
$3200$ & $8000$ & $0.8134\times 10^{-7}$ & $4.2011$  & $0.4518\times 10^{-7}$ & 4.1986 \\
$6400$ & $16000$ &  $0.4724 \times 10^{-8}$ &  $4.1058$ & $0.2633\times 10^{-8}$ & 4.1008\\
\hline
\end{tabular}
\end{table}

\begin{table}
\caption{\small\em Same as Table~\ref{T4} using the $H^1$ norm.}\label{T5}
\begin{tabular}{cccccc}
\hline
$N$ & $M$ & $\err_1[H]$ & rate for $\err_1[H]$ & $\err_1[U]$ & rate for $\err_1[U]$\\
\hline
 $200$ & $500$ &  $0.4138\times 10^{-2}$ & -- &   $0.1286 \times 10^{-1}$  & -- \\
 $400$ & $1000$ &  $0.1634\times 10^{-3}$ & $4.6624$  & $0.5075\times 10^{-3}$ & $4.6631$\\
 $800$ & $2000$  & $0.7126\times 10^{-5}$ & $4.5192$  & $0.2213\times 10^{-4}$ & $4.5189$\\
$1600$ & $4000$ &  $0.3481\times 10^{-6}$ & $4.3555$  & $0.1081\times 10^{-5}$ & $4.3549$\\
$3200$ & $8000$  & $0.1872\times 10^{-7}$ & $4.2167$  & $0.5822\times 10^{-7}$ & $4.2159$\\
$6400$ & $16000$ & $0.1082\times 10^{-8}$ & $4.1121$  & $0.3371\times 10^{-8}$ & $4.1101$\\
\hline
\end{tabular}
\end{table}

\begin{table}
\caption{\small\em Same as Table~\ref{T4} using the $L^\infty$ norm.}\label{T6}
\begin{tabular}{cccccc}
\hline
$N$ & $M$ & $\err_\infty[H]$ & rate for $\err_\infty[H]$ & $\err_\infty[U]$ & rate for $\err_\infty[U]$\\
\hline
$200$  & $500$  & $0.6037\times 10^{-2}$ &  -- &   $0.1491\times 10^{-1}$  & -- \\
$400$ & $1000$ &  $0.2386\times 10^{-3}$ & $4.6610$ &  $0.5889\times 10^{-3}$ & $4.6623$\\
$800$ & $2000$  & $0.1041\times 10^{-4}$ & $4.5176$  & $0.2569\times 10^{-4}$ & $4.5183$\\
$1600$ & $4000$ &  $0.5106\times 10^{-6}$ & $4.3507$ &  $0.1256\times 10^{-5}$ & $4.3539$\\
$3200$ & $8000$  & $0.2787\times 10^{-7}$ & $4.1953$  & $0.6777\times 10^{-7}$ & $4.2126$\\
$6400$ & $16000$ &  $0.1713\times 10^{-8}$ & $4.0239$ & $0.3960\times 10^{-8}$ & $4.0969$\\
\hline
\end{tabular}
\end{table}

\subsection{Accuracy in shape, phase, and Hamiltonian}\label{sec:propag}

The results of the spatial and temporal accuracy show that the \acs{FEM} scheme is optimally accurate in all the standard norms and also that the actual errors are very small. To further test the accuracy of this scheme, we consider the propagation of a solitary wave as in Section~\ref{sec:spatial}, while using several other norms that are pertinent to solitary waves (cf.~\cite{BDKMc}).

First, since the solitary wave's peak amplitude  remains constant during propagation, we define the normalized peak amplitude error as
\begin{equation}
\label{eq:AE}
  \err_{amp}[F] \ \doteq \ \frac{ \left| F(\xa(t),t) - F_0\right|  }{ F_0 }~,
\end{equation}
where $\xa(t)$ is the curve along which the computed approximate solution $F(x,t)$ achieves its maximum (see Section~\ref{sec:spatial}) and $F_0 \equiv \Fex (x,0)$ is the initial peak amplitude of the solitary wave. Monitoring $\err_{amp}$ as a function of propagation time, we observe that it remains very small and practically constant during propagation, \ie, $\err_{amp}[H] \approx 1.5066\times 10^{-5}$ and $\err_{amp} [U] \approx 1.2076\times 10^{-5}$. Furthermore, recall that the exact solitary wave solution travels with speed $c_s = 1.5$. We recover the solitary wave's traveling speed as 
\begin{equation}\label{eq:spped}
  \tilde{c}_s  \ \doteq \ \frac{\xa(t)-\xa(t-\tau)}{\tau}~,
\end{equation}
where $\tau$ is a constant. The results using $\tau=10$ are such that $\tilde{c}_s$ coincides with $c_s$ within the computed precision, \ie, double precision on a {\textsf{GNU Fortran}} compiler parallelized using {\textsf{OpenMP}}. This serves as additional indications of the {\bf high accuracy and non-dissipativity} of this scheme. We note that the value of $\tilde{c}_s$ depends weakly on the choice of $\tau$, which indicates a phase error. This is further studied below.

Two other error norms that are pertinent to solitary waves are the \emph{shape and phase errors}, defined below. We define the normalized shape error as the distance in $L^2$  between the computed solution at time $t = t^n$ and the family of temporally-translated exact solitary waves (with the same parameters), \ie,
\begin{equation}\label{eq:shape}
  \err_{shape}[F] \ \doteq \  \min_\tau \zeta(\tau)~,
\quad \zeta(\tau) \doteq \ \frac{  \|F(x,t^n) - \Fex(x,\tau)\| }{ \|\Fex(x,0)\| }~.
\end{equation}
The minimum in~\eqref{eq:shape} is attained at some critical $\tau=\ta(t^n)$. This, in turn, is used to define the (signed) phase error as
\begin{equation}\label{eq:phase}
  \err_{phase}[F] \ \doteq \  \ta - t^n.
\end{equation}
In order to find $\tau^*$, we use Newton's method to solve the equation $\zeta''(\tau) = 0$. The initial guess for Newton's method is chosen as $\tau^0=t^n-\Delta t$. Having computed $\tau^*$, the shape error~\eqref{eq:shape} is then
$$
\err_{shape}[F] \ = \ \zeta(\ta)~.
$$
These error norms are closely related to the \emph{orbit} of the solitary wave. Loosely speaking, they measure ``softer'' properties of the wave, which are often not well conserved using dissipative schemes, even when the schemes are accurate in all the standard norms.

Table~\ref{T7} presents the shape and phase errors as functions of propagation time, using $c_s = 1.5, \Delta x = 0.1$, $\Delta t = 0.01$. We observe that both errors remain very small. Moreover, the shape error is practically constant during the propagation.

\begin{table}
\caption{\small\em Shape and phase errors [Eqs.~\eqref{eq:shape} and~\eqref{eq:phase}] for a solitary wave as functions of propagation time.}\label{T7}
\begin{tabular}{cccccccc}
\hline
$t^n$ & $\err_{shape}[H]$ & $\err_{shape}[H]$  & $\err_{phase}[H]$ & $\err_{phase}[H]$\\
\hline
$20$ & $0.1779\times 10^{-7}$ & $0.4543\times 10^{-7}$ & $-0.6636\times 10^{-8}$ & $-0.6651\times 10^{-8}$ \\
$40$ & $0.1779\times 10^{-7}$ & $0.4544\times 10^{-7}$ & $-0.1184\times 10^{-7}$ & $-0.1187\times 10^{-7}$  \\
$60$ & $0.1779\times 10^{-7}$ & $0.4543\times 10^{-7}$ & $-0.2353\times 10^{-7}$ & $-0.2355\times 10^{-7}$  \\
$80$ & $0.1779\times 10^{-7}$ & $0.4543\times 10^{-7}$ & $-0.3002\times 10^{-7}$ & $-0.3004\times 10^{-7}$ \\
$100$ & $0.1779\times 10^{-7}$ & $0.4543\times 10^{-7}$ & $-0.2353\times 10^{-7}$ & $-0.2355\times 10^{-7}$ \\
$200$ & $0.1779\times 10^{-7}$ & $0.4543\times 10^{-7}$ & $-0.6899\times 10^{-7}$ & $-0.69018\times 10^{-7}$\\
\hline
\end{tabular}
\end{table}

Next, we test the conservation of the Hamiltonian~\eqref{eq:H} and define the corresponding normalized Hamiltonian error as
\begin{equation}\label{eq:err-H}
\err_{\H}(t^n) \ \doteq \ \left| \frac{ \H[F(x,t^n)] - \H(0) }{ \H(0)} \right|~,
\end{equation}
where $\H[\cdot] \equiv \H(t)$ denotes the energy functional~\eqref{eq:H}.

\begin{figure}
\centering
{\includegraphics[width=\columnwidth]{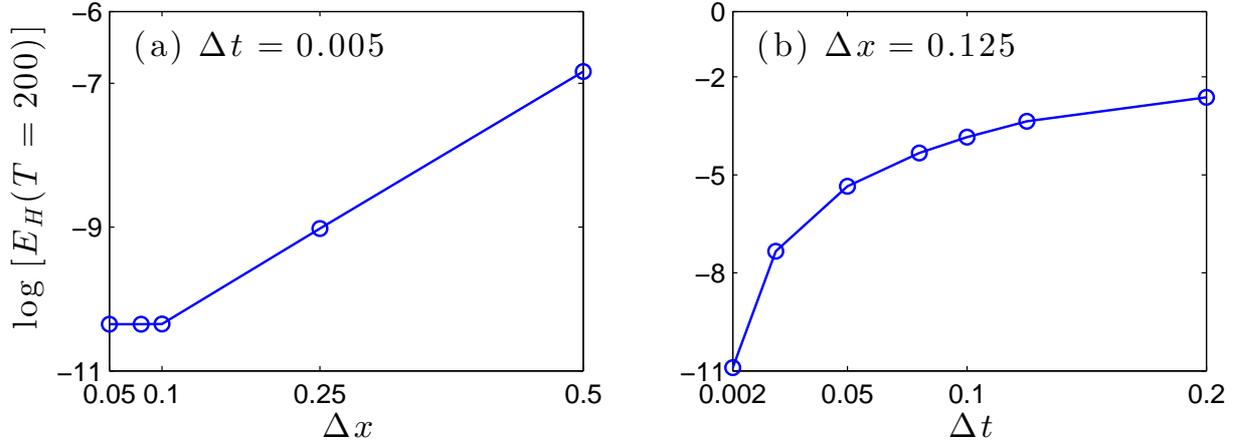}}
\caption{\small\em The log of the Hamiltonian error, $\err_{H}(T=200)$, as a function of (a) $\Delta x$ and (b) $\Delta t$.}
\label{fig:hamilton}%
\end{figure}

Table~\ref{T8} shows the results using the same wave parameters and grid sizes as in Table~\ref{T7} in the time interval $t^n\in [0,200]$. These results show that the Hamiltonian is conserved within at least 8 decimal digits of accuracy in this interval and for the specific value of $\Delta x=0.1$ and $\Delta t=0.01$. However, the error increases linearly with time. This is to be expected, since the \emph{explicit} \acs{RK} method for this problem is non-conservative. Figure \ref{fig:hamilton} presents the error in the Hamiltonian as a function of $\Delta x$ when $\Delta t=0.005$ fixed and as a function of $\Delta t$ when the value of $\Delta x=0.25$. We observe that the logarithm of the error $\err_{H}$ increases linearly with $\Delta x$ and as $\sqrt{\Delta t}$.

\begin{table}
\caption{\small\em Hamiltonian $\H$ [Eq.~\eqref{eq:H}] and corresponding error~\eqref{eq:err-H} for a solitary wave as functions of the propagation time.}
\label{T8}
\begin{tabular}{cccccccc}
\hline
$t^n$ & $\H$ & $\err_{H}(t^n)$ \\
\hline
$0$ &   $7.4266250954$ & $0.3208\times 10^{-11}$\\
$20$ & $7.4266250944$ & $0.1398\times 10^{-9}$\\
$40$ & $7.4266250933$ & $0.2828\times 10^{-9}$\\
$60$ & $7.4266250922$ & $0.4258\times 10^{-9}$\\
$80$ & $7.4266250912$ & $0.5688\times 10^{-9}$\\
$100$ &$7.4266250901$ & $0.7117\times 10^{-9}$\\
$200$ &$7.4266250848$ & $0.1427\times 10^{-8}$\\
\hline
\end{tabular}
\end{table}

\begin{figure}
\centering
{\includegraphics[width=\columnwidth]{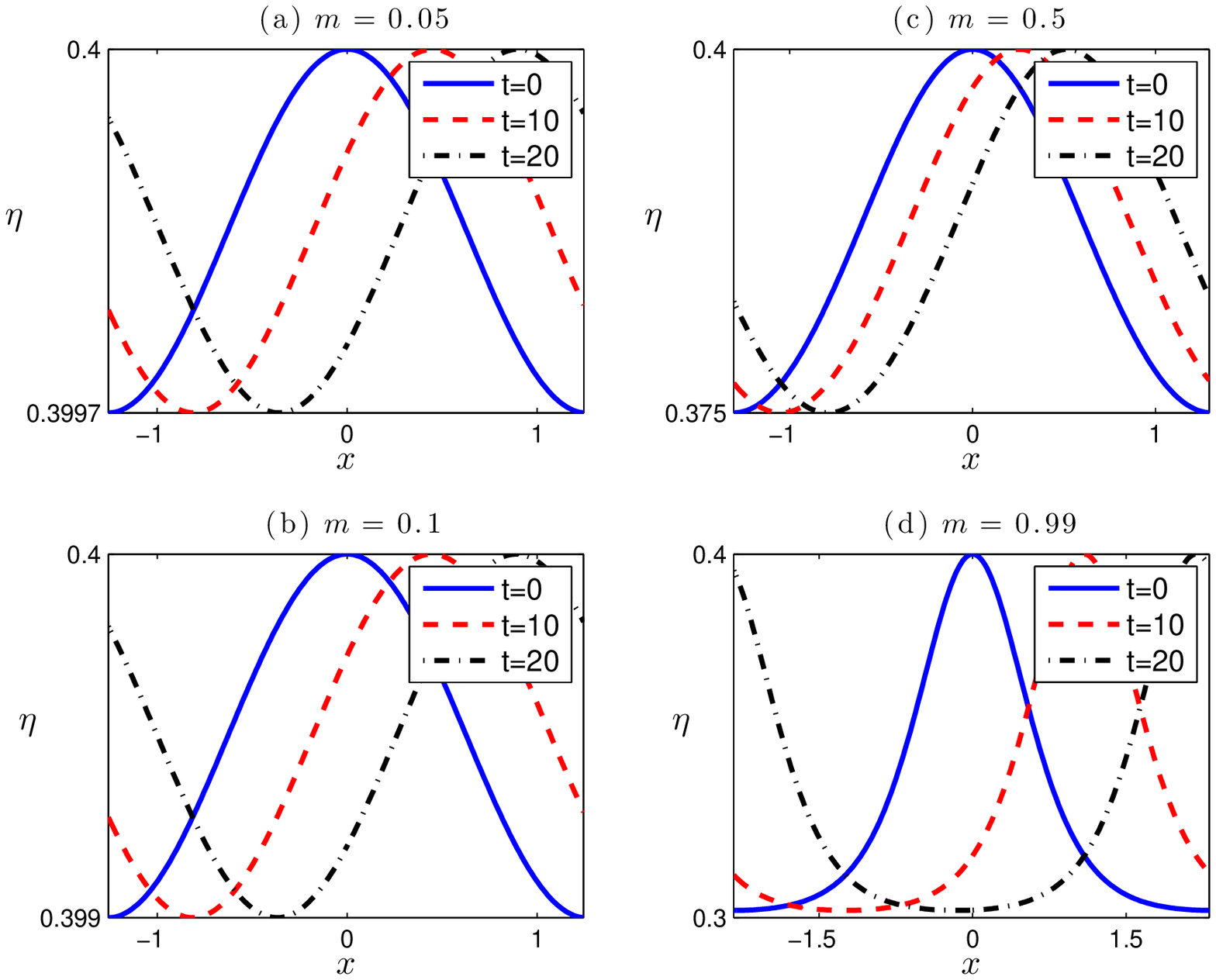}}
\caption{\small\em Propagation of cnoidal waves with $(a_0, a_1) = (0.3, 0.1)$ for four dif{}ferent values of the elliptic modulus $m$. Shown is the free surface elevation $\eta(x,t)$ at three dif{}ferent propagation times (see legends). The axes are chosen to fit a single period and the amplitude span of the waves.}
\label{F1}
\end{figure}

We close this section by computing the same errors norms for {\bf cnoidal waves}. Specifically, we consider the cnoidal wave solution~\eqref{eq:cn} with $(a_0, a_1) = (0.3, 0.1)$ and $m \in \{0.05, 0.1, 0.5, 0.99\}$. For this choice of $(a_0,a_1)$, the cnoidal waves are spectrally unstable for $m > 0.09$ (see~\cite{Carter2011}). In these computations, we consider a domain of length equal to one period, with very small values for $\Delta x$ and $\Delta t$, \ie, $N=200$ and $\Delta t=10^{-3}$. The profiles of the propagation of these cnoidal waves are presented in Figure~\ref{F1}. Table~\ref{Tcn} presents some of the error results. In all cases, the cnoidal waves propagate without significant changes in their amplitude, speed, shape, phase, and Hamiltonian. In particular, the Hamiltonian is conserved very well to within double precision. These results also show that the phase and shape errors increase as $m$ increases, especially as $m$ approaches 1. This is expected, because as $m$ increases, the cnoidal wave becomes steeper and, in the limit $m \to 1$, it approaches a solitary wave.

\begin{table}
\caption{\small\em Shape and phase errors for the computed solution $H(x, t^n)$ at $t^n = 20$, and the Hamiltonian error, $\err_{\H}$, for cnoidal waves  with $(a_0, a_1) = (0.3, 0.1)$, at four dif{}ferent values of the elliptic modulus $m$.}\label{Tcn}
\begin{tabular}{ccccc}
\hline
$m$ & $\err_{shape}[H]$ & $\err_{phase}[H]$  &  $\err_{\H}$ \\
\hline
$0.05$ & $0.1960\times 10^{-12}$ &$-0.5471\times 10^{-11}$ & $0.1005\times 10^{-16}$  \\
$0.1$ & $0.7843\times 10^{-12}$ &$-0.1314\times 10^{-10}$ & $0.2447\times 10^{-17}$  \\
$0.5$ & $0.2337\times 10^{-10}$ &$-0.9521\times 10^{-12}$ & $0.3954\times 10^{-17}$  \\
$0.99$ & $0.1637\times 10^{-8}$ &$-0.4267\times 10^{-8}$ &  $0.1268\times 10^{-16}$  \\
\hline
\end{tabular}
\end{table}

\subsection{Stability of the \acs{FEM} scheme}\label{sec:stab}

Here we perform a series of computations to study the stability of the \acs{FEM} scheme. First, we consider the propagation of a solitary wave with $c_s = 1.5$, using three different CFL ratios, \ie,
\begin{equation}
  \label{eq:cf{}l}
  \frac{\Delta t}{\Delta x} \ \in \left\{ 1, 1.5, 2, 2.1, 3 \right\}~.
\end{equation}
Table~\ref{Tstab} presents the values of the normalized shape and phase errors for the case $\Delta t = 2\Delta x$. For example, when $\Delta t = 2\Delta x$ and $\Delta x = 0.1$, the solitary wave propagates without significant changes in shape and speed. The results in the other cases are comparable except the cases where  $\Delta t > 2\Delta x$ where the solution does not remain stable. The fact that the CFL ratio can be chosen greater than $1$ indicates that the \acs{FEM} is {\bf very stable}. This is rather surprising, considering the use of an explicit \acs{RK} method.

\begin{figure}
\centering
{\includegraphics[width=0.99\columnwidth]{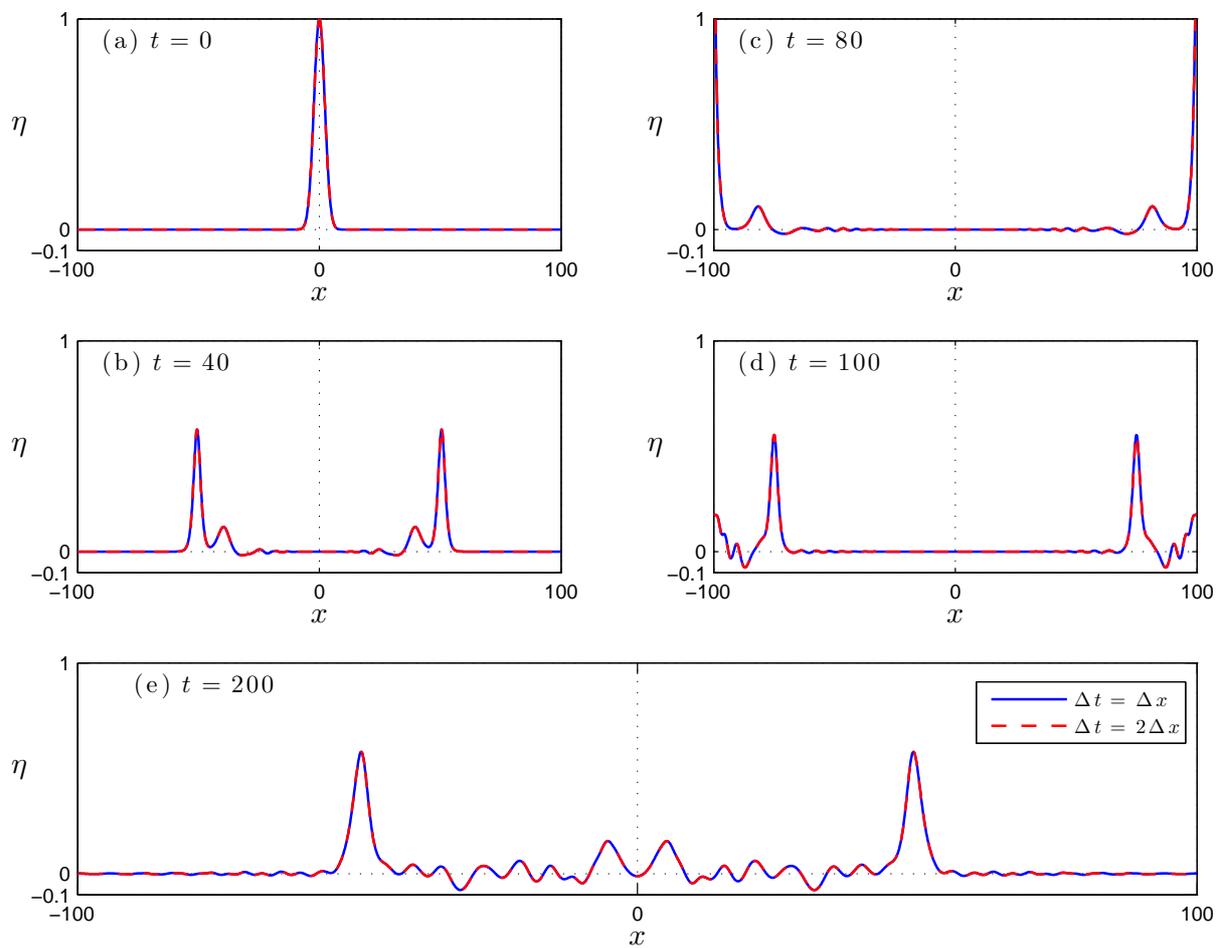}}
\caption{\small\em Breakup of a Gaussian hump [Eq.~\eqref{eq:Gaussian} with initial amplitude $A=1$ and width $\lambda=10$] into two solitary waves traveling  in opposite directions and dispersive tails. Shown are the results using $\Delta x=0.1$ and two dif{}ferent values of $\Delta t$ [see legend in (e)], which are almost indistinguishable.}
\label{F2}%
\end{figure}

\begin{figure}
\centering
{\includegraphics[width=0.99\columnwidth]{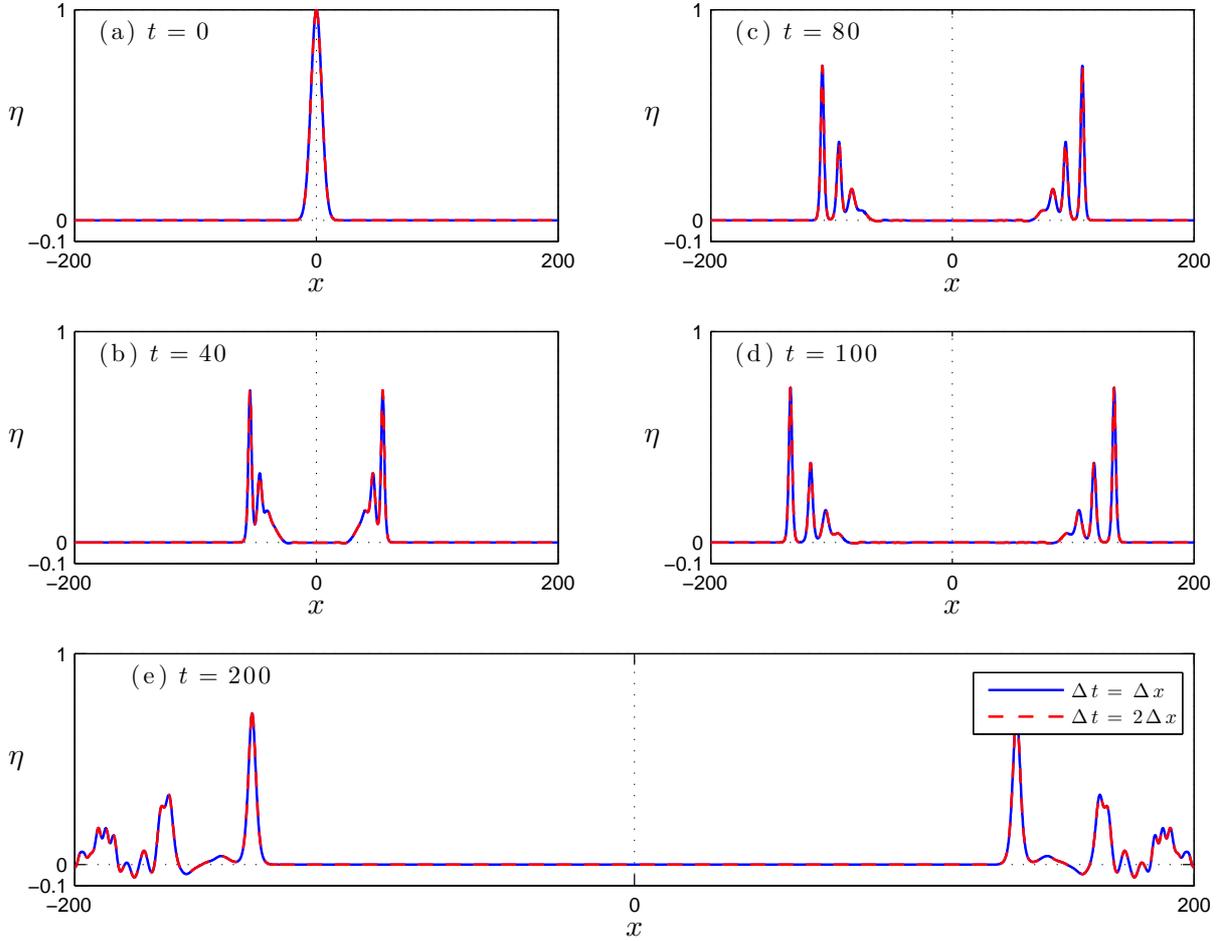}}
\caption{\small\em Same as Fig.~\ref{F2} using the initial width $\lambda=40$.}\label{F3}
\end{figure}

To further test this property, we consider  initial conditions representing a heap of water, 
\ie, 
\begin{equation}
  \label{eq:Gaussian}
\eta_0(x)=Ae^{-x^2/\lambda}, \quad u_0(x)=0~,
\end{equation}
where $A$ and $\lambda$ are constants. In all cases, the scheme is stable for large values of propagation time $t$ and all the CFL ratios in~\eqref{eq:cf{}l}. For example, Fig.~\ref{F2} presents the solutions using $A=1, \lambda=10$ and two CFL ratios: $\Delta t=\Delta x$ and $\Delta t=2\Delta x$. (For CFL$>2$ the solutions were unstable). In this case, the initial hump breaks up into two large solitary waves and smaller dispersive tails. These waves and the dispersive tails travel in opposite directions. Figure~\ref{F3} shows the results using $A = 1$ and $\lambda = 40$. Here, the solution breaks up into pairs or a larger number of solitary waves, which travel in opposite directions. In all cases, the scheme is stable even when $\Delta t = 2\Delta x$ and the difference between the solutions using the two time steps remains negligible. These results give further indication that {\bf the stability of this scheme does not impose a restrictive condition on $\Delta t$}, such that $\Delta t \le C (\Delta x)^r$ for $r > 1$.

\begin{table}
\caption{\small\em Shape and phase errors for a solitary wave as functions of propagation time, using $\Delta x=0.1$ and $\Delta t=0.2$.}
\label{Tstab}
\begin{tabular}{cccccccc}
\hline
$t^n$ & $\err_{phase}[H]$ & $\err_{phase}[H]$  & $\err_{phase}[H]$ & $\err_{phase}[H]$\\
\hline
   $ 20$ &  $0.3031\times 10^{-5}$ &  $-0.7622\times 10^{-4}$ & $0.1997\times 10^{-4}$ &  $-0.7840\times 10^{-4}$ \\
   $ 40$ &   $0.3526\times 10^{-5}$ & $-0.1787\times 10^{-3}$ & $0.2191\times 10^{-4}$ &  $-0.1810\times 10^{-3}$\\
   $ 60$ &   $0.4134\times 10^{-5}$  & $-0.3238\times 10^{-3}$ &  $0.2487\times 10^{-4}$ & $-0.3261\times 10^{-3}$\\
   $ 80$ &   $0.4815\times 10^{-5}$  & $-0.5114\times 10^{-3}$  & $0.2859\times 10^{-4}$ & $-0.5139\times 10^{-3}$\\
   $100$ &   $0.5556\times 10^{-5}$  & $-0.7423\times 10^{-3}$ &  $0.3263\times 10^{-4}$ & $-0.7445\times 10^{-3}$\\
   $200$ &   $0.9518\times 10^{-5}$  & $-0.2527\times 10^{-2}$ &  $0.5709\times 10^{-4}$ & $-0.2529\times 10^{-2}$\\
\hline
\end{tabular}
\end{table}

\section{Numerical experiments of solitary waves and DSWs}\label{sec:experiments}

In this section we present numerical experiments illustrating the behavior of the solitary waves and \acf{DSWs} in the Serre and cB system. Systems~\eqref{eq:Serre} and~\eqref{eq:cB} are solved using Algorithm~\ref{alg:Serre} and Algorithm~\ref{alg:cB}, respectively. Notwithstanding the apparent stability of both \acs{FEM} schemes, we use $\Delta t = \Delta x/10$ in order to ensure that the errors resulting from the time discretization are negligible even if the numerical solution is stable for larger values of $\Delta t$.

\subsection{Interactions of solitary waves}\label{sec:interaction}

When solitary waves interact, they incur a phase shift. In non-integrable systems, such interactions are often accompanied by the generation of small amplitude dispersive tails. Capturing this dynamics accurately requires a highly accurate scheme. Here, we study two kinds of interactions of solitary waves.
\begin{enumerate}
  \item {\bf Head-on collisions} of counter-propagating solitary waves.
  \item {\bf Overtaking collisions} of solitary waves co-propagating at dif{}ferent speeds.
\end{enumerate}

For the {\bf head-on collisions}, we generated initial conditions using~\eqref{eq:soliton} for each wave with speed $c_s=1.4$ and amplitude $A = 0.96$. The waves are initially well-separated, \ie, their peak amplitudes are located at $x = \pm 50$. The spatial domain is $x\in (-200, 200)$ and the grid sizes are
$\Delta x = 0.1$ and $\Delta t = 0.01$.

In order to make a meaningful comparison with solitary waves in the cB system, the cB solitary waves need to travel at the same speed. Since there is no known exact closed formula for  cB solitary waves, we compute them using a fixed-point iterative scheme, in which the wave's speed, $c_s$, enters as a parameter (see~\cite{Duran2013}). One upshot of this is that, for the same $c_s = 1.4$, the corresponding cB solitary wave has a somewhat larger amplitude, \ie, $A\approx 1.14763$. Moreover, the cB and~Serre waves have significantly dif{}ferent shapes.

Figures~\ref{F4} and~\ref{F5} present the solutions of the Serre and cB systems at dif{}ferent propagation times. As expected, the waves collide and emerge with small dispersive tails. Figure~\ref{F6} presents the peak amplitude, and the location of the peak amplitude as functions of time, $\xa(t)$, computed via~\eqref{eq:x_peak}. We note that, $t\mapsto\xa(t)$ is not a globally continuous function. Indeed, Fig.~\ref{F6} shows that for the Serre solution, $\xa(t)$ is discontinuous at $t\approx 38$ and $t\approx 40.5\ $. To better understand this picture, Figs.~\ref{F4} and~\ref{F5} show that, for the Serre system, as the colliding waves separate, the location of the peak amplitude changes abruptly between $t = 38$ and $t = 39$, as two of{}f-center humps grow larger than the center hump. A similar phenomenon occurs at each peak around $t = 40.5$, though with less distinguishable humps.

There are several interesting similarities and dif{}ferences between the results for the cB and~Serre systems.
\begin{itemize}
  \item[$\bullet$] The dispersive tails in the Serre system are considerably larger.
  \item[$\bullet$] During the interaction, the peak amplitude reaches approximately the same value in both systems, but at somewhat dif{}ferent times, \ie, a maximum of approximately $2.5$ at $t\approx 36.7$ for the  Serre system and a maximum of approximately $2.547$ at $t\approx 35.9$ for the cB system. Furthermore, the Serre system gives rise to the large of{}f-center humps and associated with the discontinuities in Fig.~\ref{F6}, whereas, the corresponding cB system does not have this phenomenon.
  \item[$\bullet$] The collision in the Serre system lasts longer.
  \item[$\bullet$] The change in the wave's long-time amplitude (suf{}ficiently after the collision) compared with the initial amplitude is much larger in the Serre system, \ie, it decreases by approximately $4.9\%$in the Serre system, whereas, in the cB system, it decreases by only $0.061\%$.
  \item[$\bullet$] The phase shift is significantly larger in the Serre system. The ensuing wave trajectories are closer together (with respect to linear propagation) in the Serre system.
\end{itemize}

From these experiments, we conclude that solitary waves behave qualitatively the same in the Serre and cB systems. However, quantitatively, {\bf in the interaction of the solitary waves in the Serre system is more inelastic.} These results are consistent with the fact that the Serre system contains nonlinear dispersive terms not present in the cB system. We also note that the Hamiltonian in this experiment is conserved to within 9 decimal digits and its conserved value was $\H(t) = 9.13794051$ for up to $T = 200$.

\begin{figure}
\centering
{\includegraphics[width=0.99\columnwidth]{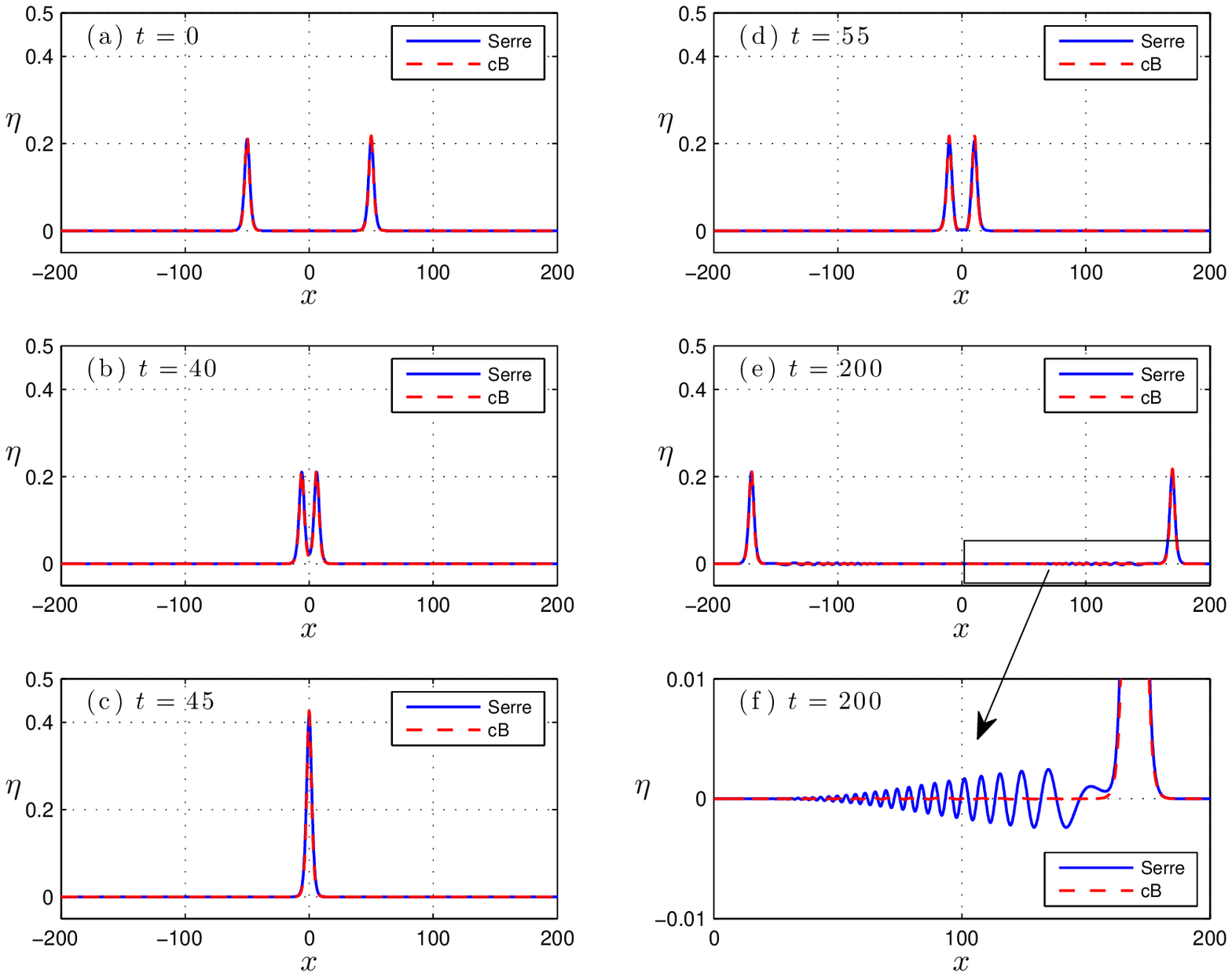}}
\caption{\small\em Same as Fig.~\ref{F3} using solitary waves with speed $c_s=1.2\,$.
(f) is a zoomed in version of (e).}\label{F7}
\end{figure}

We note that, given our choice of $\eps=1$ and solitary wave amplitude of $\O(1)$, physically speaking, these solitary waves have very large amplitudes. For this reason, cB system might not be valid in this regime. We also compare the two systems in the small-amplitude regime, in which both systems are valid, \ie, repeat the head-on collision experiments using small-amplitude solitary waves with $A\approx 0.2$ and $c_s = 1.1$. (More specifically the amplitude of the solitary waves in the case of Serre system were $A = 0.21$ and after the interaction the amplitude have been reduced to the value $A = 0.2098659$. In the case if the cB system we have $A = 0.21774185$ before the collision and $A=0.21774165$ after.). In that case, the results of the two systems are approximately the same before and after the interaction, and we refer to Figure \ref{F7} for details. It is worth noting that the dispersive tail generated by the Serre system is larger compared to the respective tail generated by the \acs{cB} system. The Hamiltonian in this experiment was $\H = 0.6764072912$ with the conserved digits shown here.

\begin{figure}
\centering
{\includegraphics[width=0.99\columnwidth]{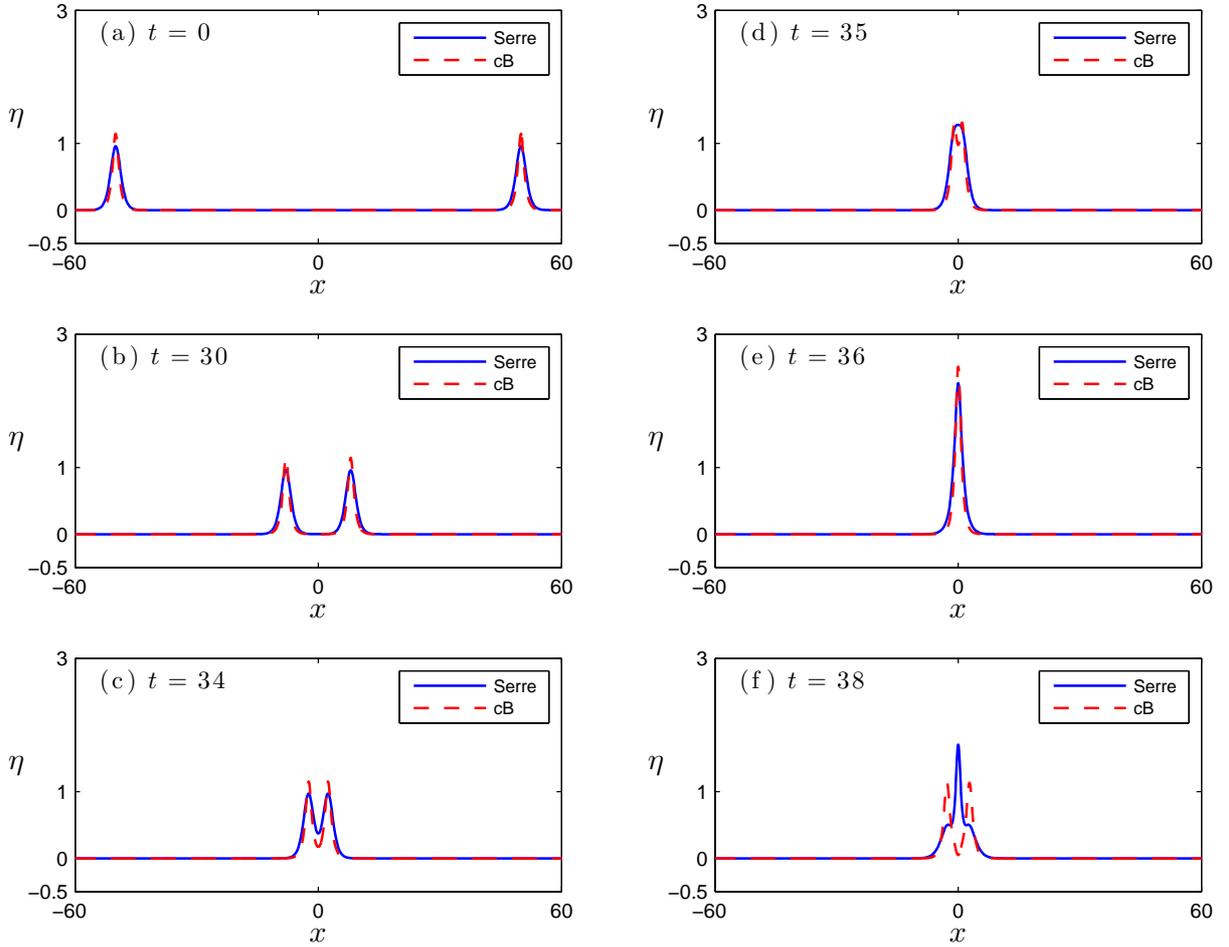}}
\caption{\small\em Head-on collision of two solitary waves in the Serre (solid) and cB (dashes) systems.}\label{F4}
\end{figure}

\begin{figure}
\centering
{\includegraphics[width=0.99\columnwidth]{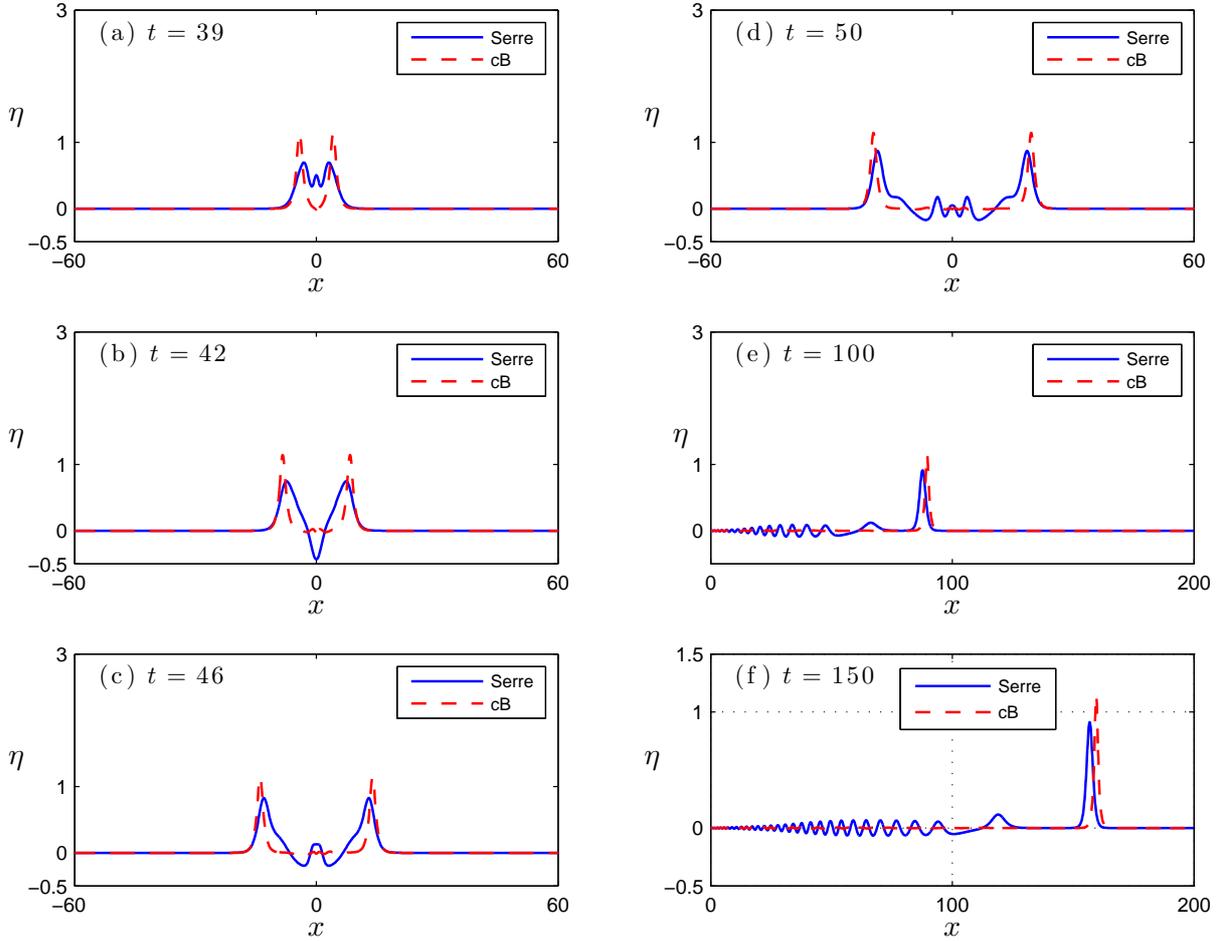}}
\caption{\small\em Continuation of the results in Fig.~\ref{F4}. Note the three humps at $t = 39$. For clarity, (e) and (f) are shown on dif{}ferent scales.}\label{F5}
\end{figure}

\begin{figure}
\centering
{\includegraphics[width=0.99\columnwidth]{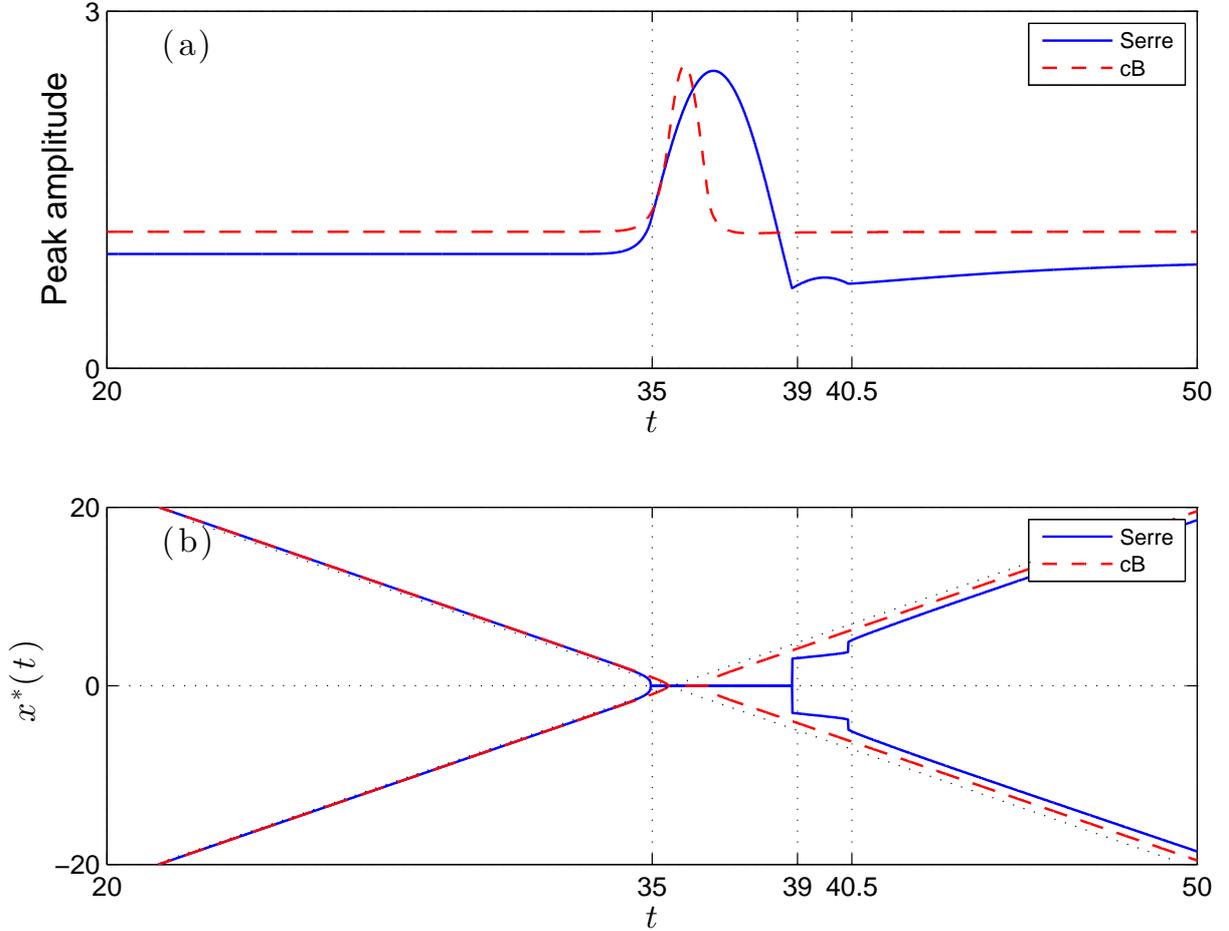}}
\caption{\small\em (a) Peak amplitude of the solution and (b) the location of the peak amplitude [see~\eqref{eq:x_peak}] of each wave for the head-on collision presented in Figs.~\ref{F4} and~\ref{F5}.}\label{F6}
\end{figure}

For the study of {\bf overtaking collisions}, we consider two solitary waves traveling in the positive $x$ direction, with speeds $c_{s,1} = 1.4$ and $c_{s,2} = 1.2$, centered at $x = \mp 50$, respectively. The amplitudes of the Serre waves are $A_1\approx 0.96$ and $A_2\approx 0.44$; and for the cB system $A_1 \approx 1.14763$ and $A_2 \approx 0.475729$.

Figures~\ref{F8}--\ref{F9} present the solutions $\eta(x,t)$ of the Serre and cB systems at dif{}ferent propagation times. Given their initial positions and speed dif{}ference, the two waves (if they were linear) should collide at $t=100/0.2 = 500$. In reality, the interaction begins at approximately $t=400$ [see Fig.~\ref{F8}(c)]. During the interaction, the waves appear to exchange mass (similar dynamics has been observed in Boussinesq systems~\cite{Dutykh2011e,AD3} and the Euler equations~\cite{CGHHS}). The faster wave overtakes the slower one at approximately $t = 490$ [see Fig.~\ref{F9}(b)]. The interaction ends at approximately $t = 600$.

A phase shift and a small change in amplitude (compared with the initial amplitudes)  is observed. Specifically, in the Serre system, the long-time amplitudes of the two waves decrease by approximately $0.025\%$ for the larger wave and $0.0705\%$ for the smaller one. In the cB system, the decrease of the amplitudes is negligible, \ie, $0.00087\%$ and $0.0027\%$, respectively. The Hamiltonian in this experiment conserved within 8 decimal digits and it was $\H = 5.7237794$ up to $T = 800$.

Furthermore, Fig.~\ref{F10} shows in detail the dispersive tails after the interaction. These tails contain $N$-shaped wavelets. The generation of wavelets has been studied recently for the cB system and other Boussinesq-like systems (cf.~\cite{AD3, ADM2}), as well as for the Euler equations (cf.~\cite{Choi1999, CGHHS}). In~\cite{Duran2013}, a related system based on a Galilean invariant equation, which contains some (but not all) of the nonlinear terms of the Serre system, showed how the wavelets depend on the nonlinear terms.

Here, Fig.~\ref{F10} shows that the \emph{signs} of the wavelets in both systems are the same, but their \emph{amplitudes} dif{}fer, \ie,  the wavelet is larger and travels faster in the Serre system. Moreover, the dispersive tail is larger in the Serre system.

These results show that:
\begin{itemize}
  \item[$\bullet$] The interaction in the Serre system is significantly stronger.
  \item[$\bullet$] Compared with the head-on collisions, the overtaking collision is significantly weaker in terms of the amplitude and phase shifts and the size of the dispersive tails.
\end{itemize}

\begin{figure}
\centering
{\includegraphics[width=0.99\columnwidth]{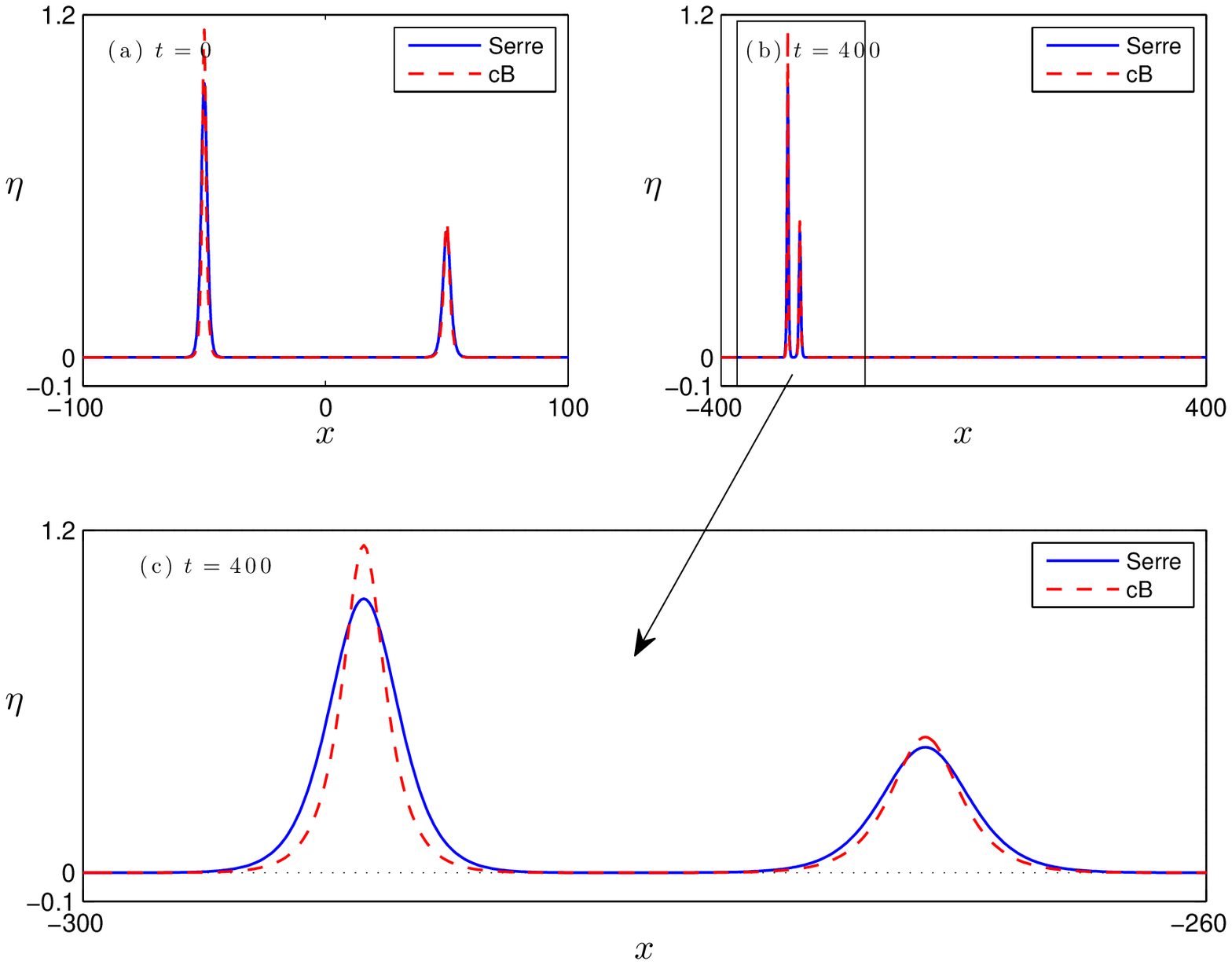}}
\caption{\small\em Overtaking collision of two solitary waves. (c) is a zoomed in version of (b), which is the beginning of the interaction.}\label{F8}
\end{figure}

\begin{figure}
\centering
{\includegraphics[width=0.99\columnwidth]{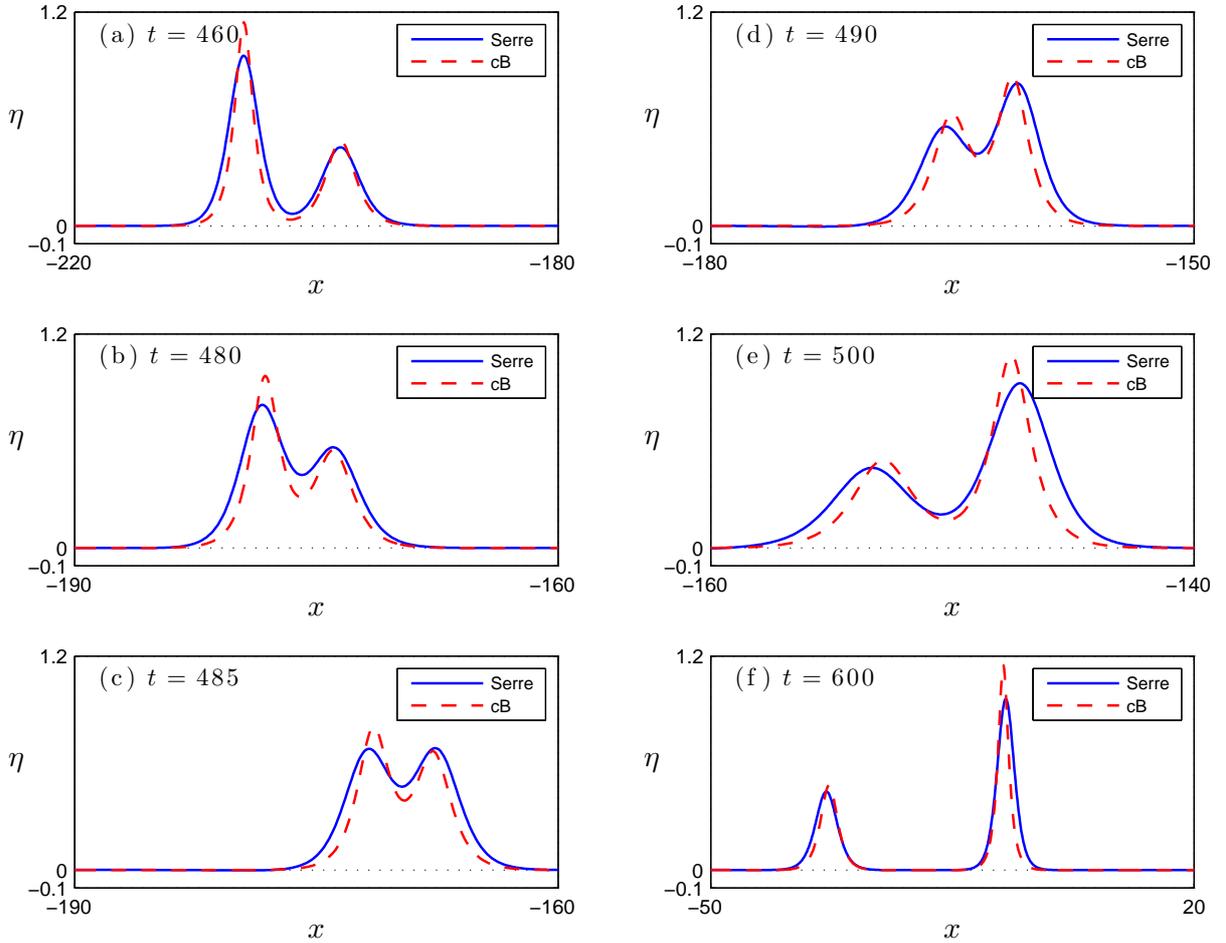}}
\caption{\small\em Continuation of the simulation in Fig.~\ref{F8}, showing the overtaking.}\label{F9}
\end{figure}

\begin{figure}
\centering
{\includegraphics[width=0.79\columnwidth]{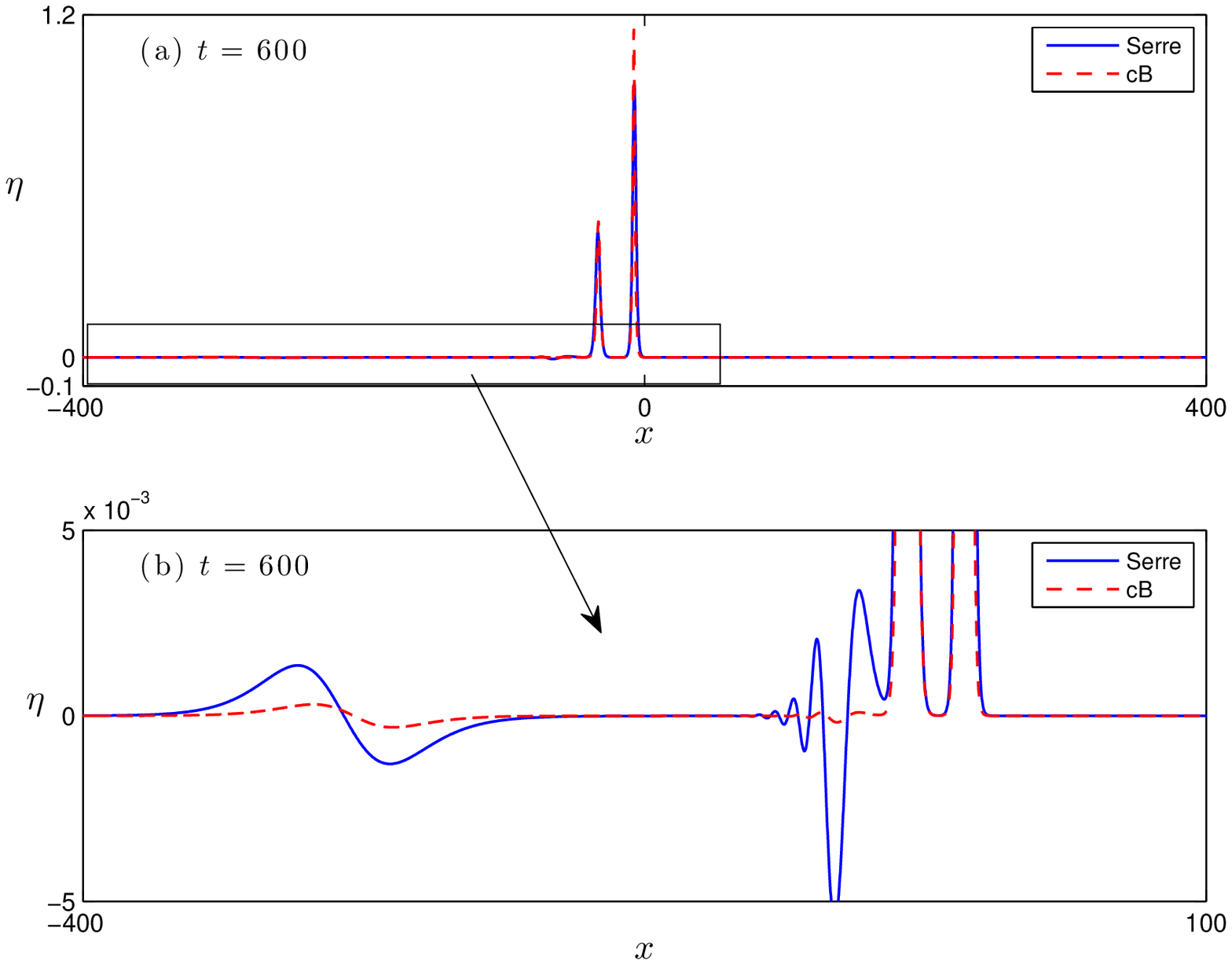}}
\caption{\small\em Same as Fig.~\ref{F9}, showing the dispersive tail in detail.}\label{F10}
\end{figure}

\subsection{Dispersive shock waves}\label{sec:DSWs}

Here we test the ability of the \acs{FEM} scheme to compute DSWs with high accuracy. In particular, we test the \acs{FEM} scheme in two cases, \ie, for a {\bf simple DSW} and for the {\bf dam break problem}. The rapid oscillations in DSWs make it challenging to simulate these problems accurately. Moreover, finite-volume and other methods are prone to adding spurious dissipative ef{}fects. This can lead to \emph{viscous-DSWs}, which look like DSWs, but travel more slowly and have smaller-amplitude  oscillations~\cite{El2006}. One of the advantages of the \acs{FEM} scheme is that it is non-dissipative, as shown below.

First, we study {\bf simple DSWs} in the Serre and cB systems. The simulations are carried on the interval $x\in (-700, 700)$ with $\Delta x = 0.1$ and $\Delta t = 0.01$. We choose as initial data for $h$ as a step function that decays to zero as $|x| \to \infty$, \ie, 
\begin{equation}\label{eq:h0}
h(x,0) \ =\ \frac{1}{2} \eta_0 \left [ 1+\tanh \left(250-|x| \right) \right], 
\end{equation}
where $\eta_0=0.4182\ $. The initial data for $u(x,0)$ is chosen as
$$
  u(x,0) \ =\  2\left[\sqrt{h(x,0)}-1\right]~.
$$
These initial data generate a simple DSW with (see~\eqref{eq:u-} and~\cite{El2006, El2008})
\begin{equation}\label{eq:f{}low-asymp}
  h^+=1, \quad h^- = 1+\eta_0, \quad u^+=0, \quad u^- = 2(\sqrt{1+\eta_0}-1)~.
\end{equation}

\begin{figure}
\centering
\includegraphics[width=0.79\textwidth]{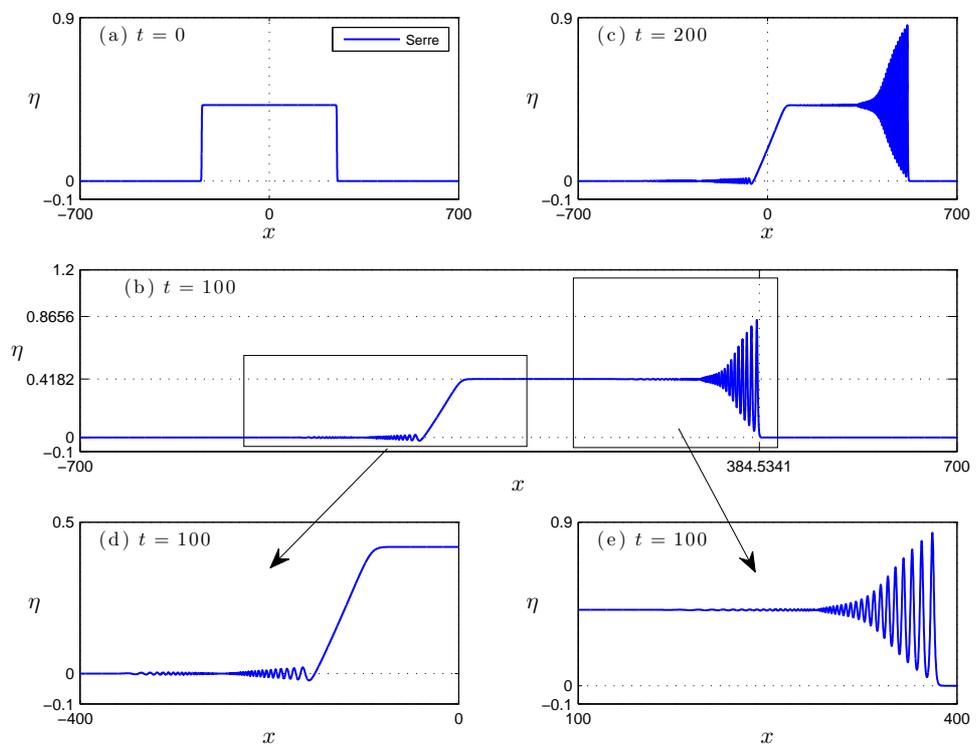}
\caption{\small\em Simple DSW in the Serre system.}\label{F11}
\end{figure}

\begin{figure}
\centering
\includegraphics[width=0.99\textwidth]{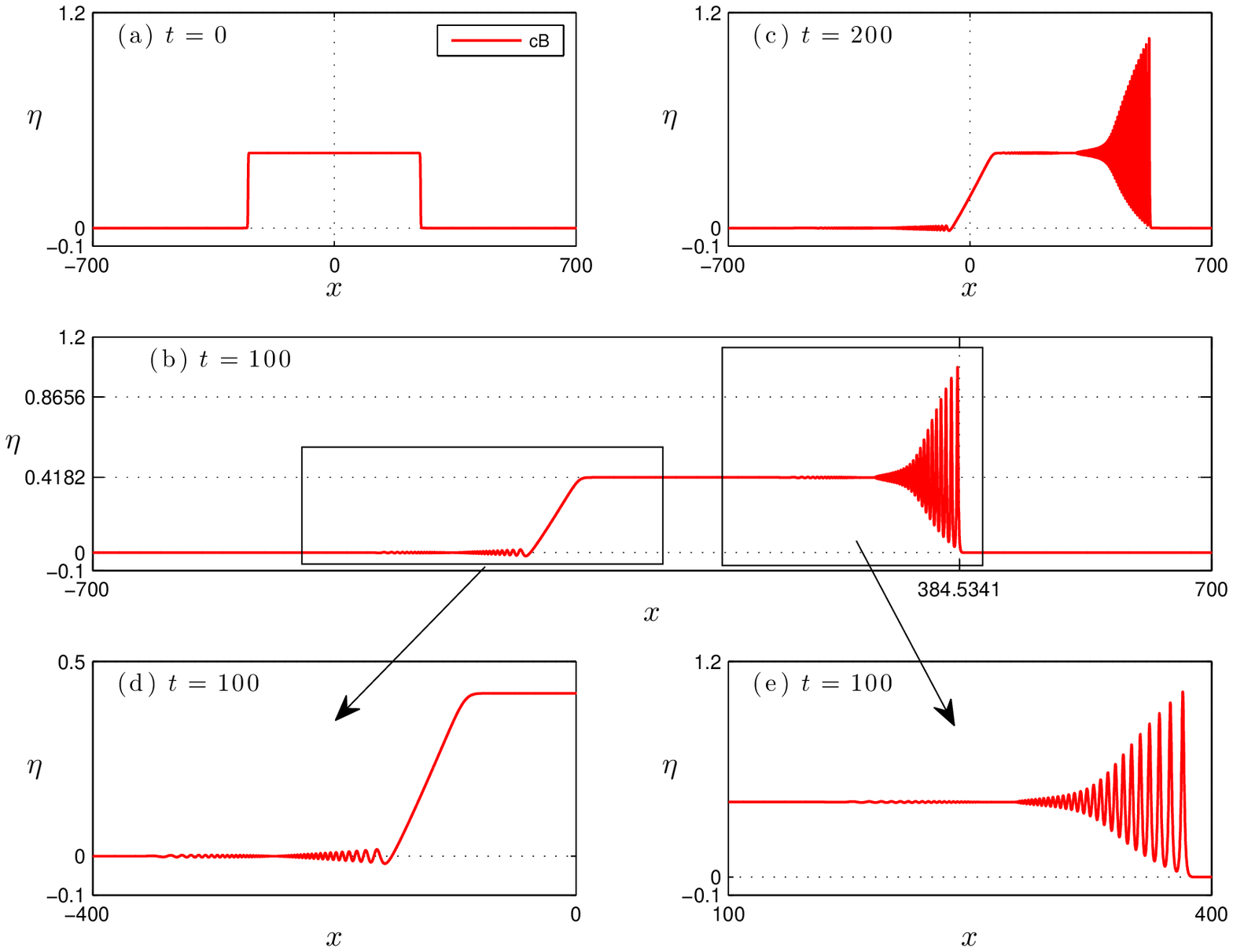}
\caption{\small\em Same as Fig.~\ref{F11} in the cB system.}\label{F12}
\end{figure}

Figures~\ref{F11} and~\ref{F12} show the results for the Serre and cB systems, respectively. In both systems, a simple DSW is generated, which travels to the right, and a rarefaction wave travels to the left with a small dispersive tail.

\begin{figure}
\centering
\includegraphics[width=0.99\textwidth]{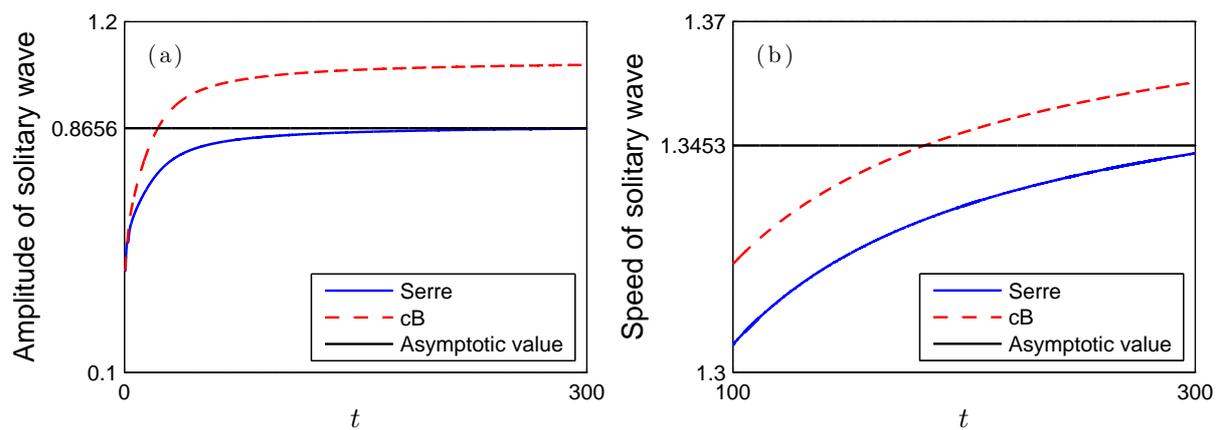}
\caption{\small\em Amplitude and speed of the leading-edge solitary wave for the simulations in Figs.~\ref{F11} and~\ref{F12}. Also shown are the asymptotic values for the Serre system [solid horizontal lines, Eqs.~\eqref{eq:DSW-simple}].}\label{F13}
\end{figure}

To test the non-dissipativity of the \acs{FEM} scheme, we compare the computational results with the asymptotics of the leading edge solitary wave, whose long-time amplitude and speed are given in Eqs.~\eqref{eq:DSW-simple} with the jump [Eq.~\eqref{eq:delta}] $\delta = \eta_0 = 0.4182$. Here, $a_1^{\rm simple} \approx 0.8656$ and $c_s^{\rm simple} \approx 1.3453$. Figure~\ref{F13} shows the peak amplitude and speed of the solitary wave recovered from the computations approach the corresponding asymptotic values (solid horizontal lines). Even though $\delta$ is not much smaller than $1$, it turns out that the asymptotic values are fairly accurate. These results show that {\bf the \acs{FEM} is non-dissipative even for DSWs.} We note that the Hamiltonian in these simulations is conserved to within $10$ decimal digits of accuracy and it remained $\H(t) = 190.4720453$ even after the interaction of the leading edge with the other parts of the solution and up to $T=400$. In addition, Fig.~\ref{F13} shows that, in the cB system, the DSW travels significantly more slowly and with a larger amplitude.

\begin{figure}
\centering
\includegraphics[width=0.79\textwidth]{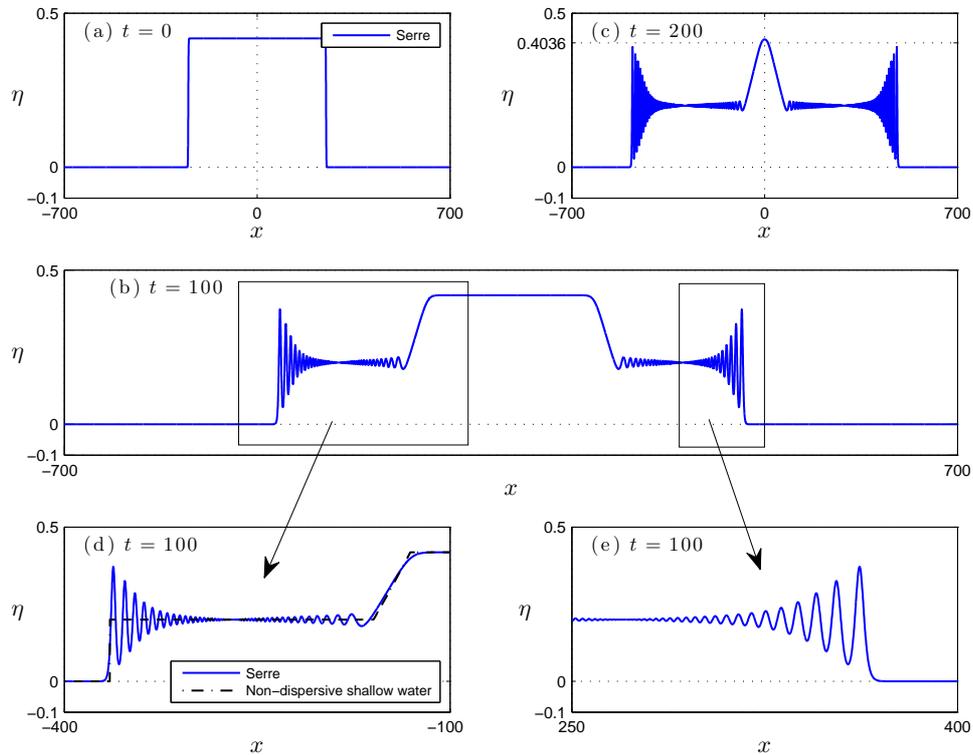}
\caption{\small\em Dam break in the Serre system. Part (d) also shows the shock and rarefaction waves for the corresponding non-dispersive shallow water problem (dot-dashes).}\label{F14}
\end{figure}

\begin{figure}
\centering
\includegraphics[width=0.99\textwidth]{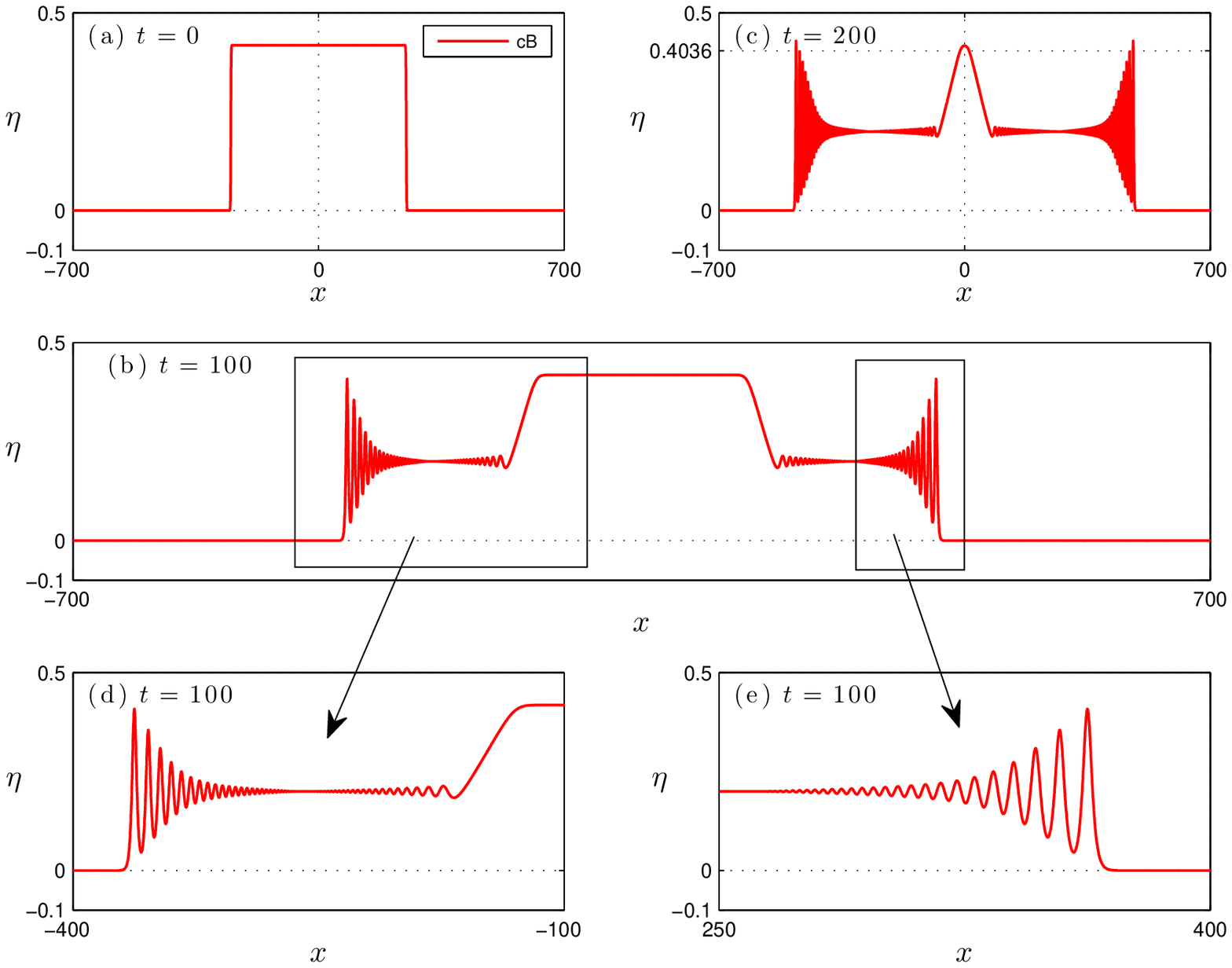}
\caption{\small\em Same as Fig.~\ref{F14} in the cB system.}\label{F15}
\end{figure}

\begin{figure}
\centering
\includegraphics[width=0.79\textwidth]{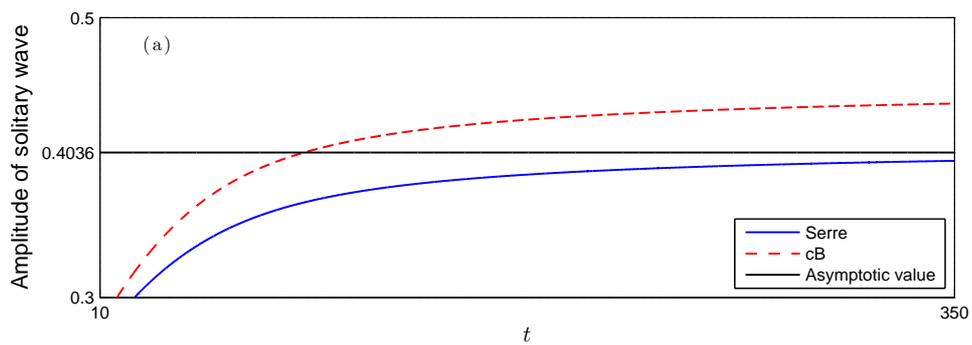}
\caption{\small\em Amplitude of the leading-edge solitary waves for the simulations in Figs.~\ref{F14} and~\ref{F15}. Also shown is the asymptotic value for the Serre system [solid horizontal lines, Eq.~\eqref{eq:DSW-dam-a1}].}\label{F16}
\end{figure}

We also consider the {\bf dam-break problem} (see Section~\ref{sec:DSW}). Here, the initial data for $h(x,0)$ are the same as~\eqref{eq:h0}, but  $u(x,0) = 0$. Figure~\ref{F14} shows the results of this computation for the Serre system, \ie, two counter-propagating DSWs and rarefaction waves. These initial data generate a simple DSW with~\eqref{eq:f{}low-asymp}, whose leading edge solitary wave has amplitude and speed given by~\eqref{eq:DSW-dam} with $\delta = \eta_0  = 0.4182$. For comparison, Fig.~\ref{F14}(d) shows the corresponding \emph{non-dispersive} shallow water shock and rarefaction waves, which connect the same f{}low states. Similar results are obtained for the cB system using the same initial conditions -- see Figs.~\ref{F15} and~\ref{F16}. The discretization parameters are the same as in the previous experiment and the Hamiltonian is conserved with $10$ decimal digits of accuracy and it was $\H(t) = 87.27888421$.

\begin{figure}
\centering
\includegraphics[width=0.99\textwidth]{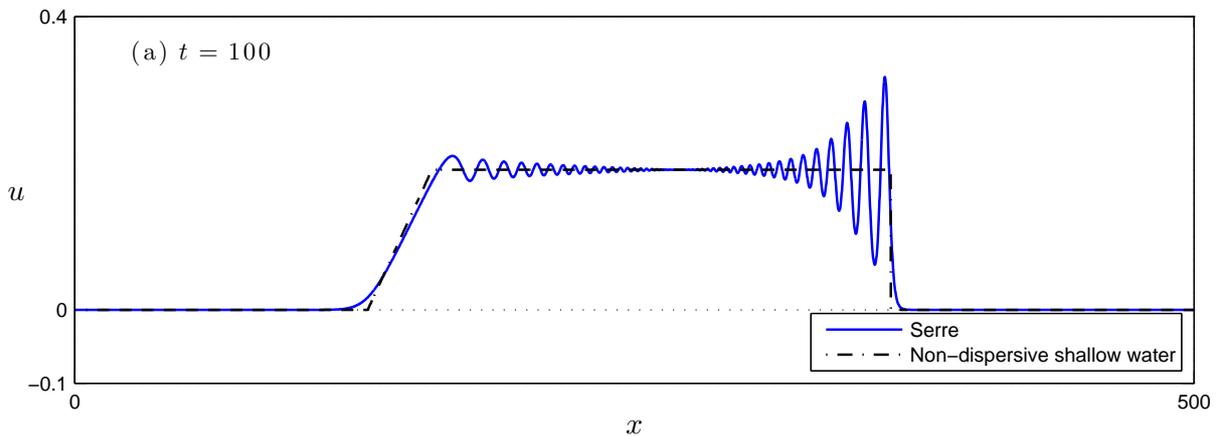}
\caption{\small\em Velocity profile of the  DSW and rarefaction waves corresponding to the free surface elevation shown in the right side of Fig.~\ref{F14}(b). Also shown are the shock and rarefaction waves for the corresponding non-dispersive shallow water problem (dot-dashes).}\label{F14b}
\end{figure}

\section{Summary and conclusions}

We present a fully discrete numerical scheme for the Serre system based on the standard Galerkin / finite-element method with smooth periodic splines and on the fourth-order, four-stage, explicit Runge--Kutta method. The computational results show that this numerical scheme is highly accurate and stable. In particular,  this scheme achieves the optimal orders of convergence in time and space. Moreover, the actual numerical errors remain fairly small during propagation. In addition, the stability of this scheme does not impose restrictive conditions on the temporal step size, suggesting that this scheme could be unconditionally stable.

In addition, we perform a series of highly-accurate numerical experiments of interacting solitary waves in the Serre and `classical' Boussinesq systems. The computational results show that the interactions of solitary waves in the Serre system are more inelastic, \ie, the interaction is significantly longer and incurrs a larger amplitude change and larger phase shift. This greater ``inelasticity'' does not af{}fect the nonlinear stability of the solitary waves. Furthermore, in the Serre system, the dispersive tails generated by the interacting solitary waves have larger amplitude.

We also use this scheme to study the generation and propagation of rapidly oscillating dispersive shocks and rarefaction waves. The results show that this scheme can resolve the fine details of the solutions, without inducing numerical (artificial) dissipative ef{}fects.

\subsection*{Acknowledgments}
\addcontentsline{toc}{section}{Acknowledgments}

D.~\textsc{Dutykh} would like to acknowledge the hospitality of UC Merced during his visit in April 2013 and the support from ERC under the research project ERC-2011-AdG 290562-MULTIWAVE. D. Mitsotakis would like to thank Prof. Mark Hoefer for his suggestions and his comments and for the fruitful discussions on dispersive waves.

\addcontentsline{toc}{section}{References}
\bibliographystyle{abbrv}
\bibliography{biblio}

\end{document}